\documentclass[iop, apj]{emulateapj}

\newcommand{\kms}{\rm km~s^{-1}}
\newcommand{\kmsmpc}{\rm km~s^{-1}~Mpc^{-1}}

\slugcomment{\bf Last updated:\today}

\usepackage{color}
\usepackage{multirow}
\usepackage{lscape} 

\begin{document}

\title{The HectoMAP Cluster Survey - II. X-ray Clusters}

\author{Jubee Sohn$^{1}$, 
        Gayoung Chon$^{2}$, 
        Hans B{\"o}hringer$^{2}$, 
        Margaret J. Geller$^{1}$,  
        Antonaldo Diaferio$^{3, 4}$, 
        Ho Seong Hwang$^{5}$, 
        Yousuke Utsumi$^{6}$,
        Kenneth J. Rines$^{7}$}
        
\affil{$^{1}$ Smithsonian Astrophysical Observatory, 60 Garden Street, Cambridge, MA 02138, USA}
\affil{$^{2}$ Max-Planck-Institut f{\"u}r extraterrestrische Physik, D-85748 Garching, Germany}
\affil{$^{3}$ Universit{\`a} di Torino, Dipartimento di Fisica, Torino, Italy}
\affil{$^{4}$ Istituto Nazionale di Fisica Nucleare (INFN), Sezione di Torino, Torino, Italy}
\affil{$^{5}$ Quantum Universe Center, Korea Institute for Advanced Study, 85 Hoegiro, Dongdaemun-gu, Seoul 02455, Korea}        
\affil{$^{6}$ Kavli Institute for Particle Astrophysics and Cosmology, SLAC National Accelerator Laboratory, Stanford University, SLAC, 2575 Sand Hill Road, M/S 29, Menlo Park, CA  94025, USA}
\affil{$^{7}$ Department of Physics \& Astronomy, Western Washington University, Bellingham, WA 98225, USA}        

%=============================================================
\begin{abstract}
We apply a friends-of-friends algorithm to the HectoMAP redshift survey 
 and cross-identify associated X-ray emission in the ROSAT All-Sky Survey data (RASS). 
The resulting flux limited catalog of X-ray cluster survey is complete to a limiting flux of  
 $\sim 3 \times 10^{-13}$ erg s$^{-1}$ cm$^{-2}$ and 
 includes 15 clusters (7 newly discovered) with redshift $z \leq 0.4$.
HectoMAP is a dense survey ($\sim 1200$ galaxies deg$^{-2}$) 
 that provides $\sim 50$ members (median) in each X-ray cluster. 
We provide redshifts for the 1036 cluster members. 
{\it Subaru}/Hyper Suprime-Cam imaging covers 
 three of the X-ray systems and 
 confirms that they are impressive clusters.
The HectoMAP X-ray clusters have an $L_{X} - \sigma_{cl}$ scaling relation
 similar to that of known massive X-ray clusters. 
The HectoMAP X-ray cluster sample predicts
 $\sim12000 \pm 3000$ detectable X-ray clusters in the RASS to the limiting flux,
 comparable with previous estimates.
\end{abstract}
%=============================================================
\section{INTRODUCTION}

Searching for clusters of galaxies is a stepping stone 
 toward understanding galaxy evolution in dense environments 
 (e.g. \citealp{Dressler84, Blanton09, Wetzel14, Haines15}), 
 and the formation of large scale structure (e.g. \citealp{Bahcall88, Postman92, Reiprich02, Chon13}), and 
 for evaluating the cosmological parameters (e.g. \citealp{Voit05, Vikhlinin09, Allen11, Bohringer14}). 
A large cluster catalog provides 
 a sample for statistical studies of the formation and evolution of galaxies 
 within dense, massive gravitationally bound systems. 
Simultaneously,  
 the number density and mass distribution of the ensemble of galaxy clusters 
 are important probes of cosmological models. 

A wide variety of techniques yield cluster catalogs.  
Initial systematic surveys for clusters 
 identified over-densities of galaxies on the sky
 \citep{Abell58, Abell89, Zwicky68}. 
Many recent studies use the red-sequence 
 that generally characterizes cluster galaxies along with the brightest cluster galaxies (BCGs)
 to identify clusters \citep{Gladders00, Koester07, Hao10, Oguri14, Rykoff14, Oguri17}.
Often the redshifts for these large photometric samples are photometric \citep{Wen09, Wen12, Szabo11, Durret15}. 
These catalogs contain a large fraction of real systems, 
 but they cannot discriminate completely against line-of-sight superpositions.
Cluster identification based on spectroscopic redshifts resolves the contamination issue, 
 but the density of a spectroscopic survey can be a limiting factor in evaluating the completeness and purity of the catalog. 

Large galaxy cluster catalogs 
 are also based on X-ray identification. 
The ROSAT All-Sky Survey (RASS) data have been especially important
 \citep{Ebeling98, Bohringer00, Bohringer04, Ebeling10, Bohringer13, Bohringer17}. 
The Sunyaev-Zel'dovich (SZ) effect \citep{Sunyaev72} is 
 a more recent but equally powerful tool
 for identifying clusters \citep{Vanderlinde10, Marriage11, Planck15, Planck16, Bleem15}. 
For verification of both the X-ray and SZ candidates, 
 optical counterparts identified from imaging and/or spectroscopy are critical. 
Optical follow-up observations for Northern ROSAT All sky survey (NORAS)
 find optical clusters associated with 76\% of the X-ray extended sources \citep{Bohringer00}. 
Similarly, \citet{Bleem15} show that
 $76\%$ of the South Pole Telescope (SPT)-SZ clusters with $4.5\sigma$ detection 
 are confirmed by optical and/or near-infrared imaging.
  
Cluster catalogs are also constructed from complete spectroscopic surveys
 (e.g. \citealp{Geller83, Mahdavi00, Finoguenov09, Robotham11, Tempel16}). 
\citet{Mahdavi00} identified galaxy groups based on a complete spectroscopic survey. 
After identification of the spectroscopic candidates, 
 they searched the RASS for these objects. 
This cross-identification identified the most reliable systems. 
More recently, 
 \citet{Starikova14} followed a similar procedure 
 where clusters were first identified with a combination of spectroscopic and
 weak lensing observations and followed with X-ray observations. 
Combining a spectroscopic survey with other cluster identification techniques is a powerful approach 
 because it provides direct cluster membership with little line-of-sight contamination. 
Furthermore the spectroscopic redshifts enable  
 an estimate of the dynamical mass of the systems for direct comparison with X-ray, SZ, and weak lensing measurements. 

Here we follow an approach similar to the one pioneered by \citet{Mahdavi00}. 
 We explore a much deeper redshift survey to a limiting $r = 21.3$.
We use the HectoMAP \citep{Geller11, Geller15} redshift survey to identify massive candidate systems 
 based on a friends-of-friends (FoF) algorithm. 
We then use the ROSAT all-sky survey 
 to search for X-ray emission associated with the clusters in the HectoMAP region 
 and to test whether the X-ray emission is consistent with  emission
 from a hot thermal intracluster plasma. 
We also check the HectoMAP cluster catalog 
 against previously published X-ray cluster candidates. 
We test our approach by applying the identical technique to 
 the SHELS redshift survey \citep{Geller10, Geller12, Geller14} explored by \citet{Starikova14}.
 
HectoMAP is a dense redshift survey complete to $r = 21.3$ 
 and with a median redshift of $z \sim 0.39$. 
The red-selected HectoMAP survey is 95\% complete in the selected color range,
 $(g-r)_{fiber, 0} > 1.0$ and $(r-i)_{fiber, 0} > 0.5$. 
Using this survey
 \citet{Hwang16} examine large-scale structures and voids 
 in comparison with the result from the Horizon Run 4 N-body simulation \citep{Kim15}. 
The observed richness and size distributions of both over- and under-dense structures
 agree well with the simulations.
  
The catalog of HectoMAP galaxy clusters described here depends on
 cross-identification with the X-ray. 
We apply the FoF to the complete color-selected galaxy catalog. 
We refine the FoF cluster membership by applying the caustic technique to all of the data in the HectoMAP region.
In the catalog, 
 we identify 15 robust clusters with a median of $\sim 50$ spectroscopically identified members. 
Seven of these clusters have not been previously identified.
We examine their physical properties 
 including X-ray luminosities and velocity dispersions. 
These detections suggest that 
 even to the limit of the RASS 
 there are more than $\sim12000$ detectable massive X-ray clusters
 to a flux limit of $\sim 3 - 5 \times 10^{-13}$ erg s$^{-1}$ cm$^{-2}$ in the 0.1 to 2.4 keV energy band 
 compatible with the estimate by \citet{Schuecker04}.
These clusters are typically at redshifts less than $0.4$. 
This sample hiding in the existing RASS data is larger than the SPT samples \citep{Bleem15}.
 
We describe the spectroscopic data from HectoMAP 
 and X-ray data from the RASS in Section \ref{data}. 
Section \ref{fof} explains the cross-identification techniques
 including the description of the cluster finding methods. 
The cluster catalog is in Section \ref{cat}. 
We also include redshifts of the 1036 cluster members. 
We discuss the results including the physical properties of the clusters in Section \ref{discussion}
 and summarize in Section \ref{summary}. 
We adopt the standard $\Lambda$CDM cosmology of 
 $H_{0} = 70~\kmsmpc$, $\Omega_{m} = 0.3$, and $\Omega_{\Lambda} = 0.7$
 throughout the paper. 

%=============================================================
\section{DATA}\label{data}

Selection of clusters from a redshift survey is 
 only possible when the sampling density is sufficiently high 
 (e.g. CNOC survey, \citealp{Yee96}).
HectoMAP has a density of $\sim 2000$ galaxies deg$^{-2}$ to a limit $r_{petro, 0} = 21.3$; 
 the median depth of the redshift survey is 0.39. 
Our approach here is to identify cluster candidates by applying an FoF to the redshift survey and 
 then using the resultant catalog as a finding list for extended X-ray sources in the ROSAT all-sky survey (RASS).

We first describe the salient features of HectoMAP (Section \ref{hmap}). 
Then we review the RASS detection limits relevant 
 for testing the HectoMAP candidate clusters against the X-ray data (Section \ref{rosat}).  

\subsection{HectoMAP: A Dense Spectroscopic Survey}\label{hmap}

HectoMAP is a dense redshift survey of red galaxies
 covering 52.97 deg$^{2}$ of the sky with $200 < $ R.A. (deg) $< 250$ and $42.5 <$ Decl. (deg) $< 44.0$
 \citep{Geller11, Geller15, Hwang16}. 
We select HectoMAP galaxies from the Sloan Digital Sky Survey (SDSS) Data Release 9 (DR9)
 \citep{Ahn12}.
We select red galaxies with $(g-r)_{fiber, 0} > 1.0$, $(r-i)_{fiber, 0} > 0.5$
 as redshift survey targets.  
The color selection removes objects with $z \lesssim 0.2$ 
 where the SDSS spectroscopic survey has reasonable coverage.
The targets have $r_{petro, 0} < 21.3$ and $r_{fiber, 0} < 22.0$. 
We fill fibers we cannot allocate to the main HectoMAP red targets with objects bluer or fainter than the target limits.
We exclude $r_{fiber, 0} > 22.0$ objects 
 because their low surface brightness makes the acquisition of a redshift difficult 
 in the standard HectoMAP exposure time. 

\begin{figure}
\centering
\includegraphics[scale=0.4]{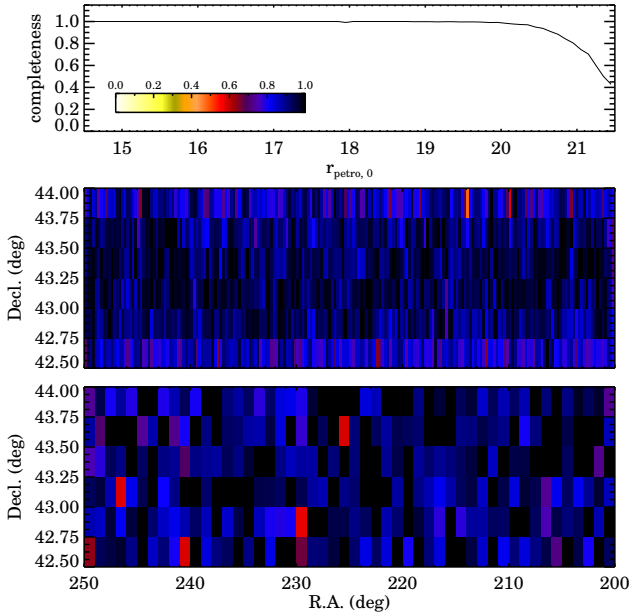}
\caption{(Top) The differential spectroscopic completeness for HectoMAP 
 as a function of $r-$ band magnitude.
(Middle) Two-dimensional completeness map ($200 \times 6$ pixels) of HectoMAP
 for the galaxies with $r_{petro, 0} \leq 21.3$,
 $(g-r)_{fiber, 0} > 1.0$, and $(r-i)_{fiber, 0} > 0.5$.
(Bottom) Same as the middle panel, 
 but for SDSS galaxies with $r_{petro, 0} \leq 17.77$
 with larger pixels ($25 \times 6$ pixels).  }
\label{complete}
\end{figure}

From 2009 to 2016, 
 we conducted a redshift survey with the 300-fiber spectrograph Hectospec 
 mounted on MMT 6.5m telescope \citep{Fabricant98, Fabricant05}.
The fiber-fed spectrograph Hectospec typically obtains spectra for $\sim250$ targets 
 within a $\sim1$ deg$^{2}$ field of view in a single observation. 
The 270 line mm$^{-1}$ grating  
 yields a wavelength range $3700 - 9150$ \AA~ with 
 a resolution of $\sim6.2$ \AA. 
The exposure time for each field is $0.75 - 1.5$ hr and 
 each exposure is comprised of three subexposures for cosmic ray removal. 
To guarantee uniform sampling even in the densest regions,
 we revisit each HectoMAP position $\sim 7$ times. 
The completeness map (Figure \ref{complete}) shows the resulting uniformity of the survey. 
The only residual non-uniformity occurs toward the edges of the field. 

We also compile previously measured redshifts in the HectoMAP region from SDSS DR14 \citep{Abolfathi17} and
 from the NASA Extragalactic Database (NED). 
There are 2143 redshifts from SDSS and 161 redshifts from the literature (NED) within the HectoMAP red selection. 
The typical uncertainties in the redshifts from SDSS and NED are $28~\kms$ and $60~\kms$, respectively. 
There are no significant zero-point offsets between the SDSS and Hectospec redshifts \citep{Geller14}. 

We reduce the Hectospec spectra using the HSRED v2.0 package 
 originally developed by Richard Cool and 
 substantially revised by the SAO Telescope Data Center (TDC) staff\footnote{http://www.mmto.org/node/536}.
The TDC has tested this reduction against 
 the original HSRED and 
 IRAF SPECROAD packages \footnote{http://tdc-www.harvard.edu/instruments/hectospec/specroad.html}. 
We derive redshifts using RVSAO \citep{Kurtz98}, a cross-correlation code. 
The set of spectral templates is identical to the set used for earlier reductions of Hectospec data. 
We visually inspect each spectrum and classify redshifts 
 into three groups: 
 `Q' for high-quality spectra, `?' for ambiguous fits, and `X' for poor fits. 
We use only `Q' spectra. 
There are 58211 redshifts for red galaxies satisfying the HectoMAP magnitude and color selection. 
The typical redshift uncertainty normalized by $(1+z)$ is $\sim 32 ~\kms$. 

% ======================================
% Figure 1 - Figure \ref{complete}
% ======================================
% ======================================

Figure \ref{complete} shows the spectroscopic completeness of HectoMAP to $r_{petro, 0} = 21.3$.
\citet{Hwang16} displayed a similar plot for the bright ($r_{petro, 0} < 20.5$) portion of HectoMAP 
 which was 89\% complete to this limit at that time. 
The survey is now 98\% complete to $r_{petro, 0} = 20.5$. 

The upper panel of Figure \ref{complete} is the current spectroscopic completeness for red galaxies 
 with $(g-r)_{fiber, 0} > 1.0$ and $(r-i)_{fiber, 0} > 0.5$ 
 as a function of apparent magnitude. 
The integral completeness of the survey to $r_{petro, 0} = 21.3$ for the HectoMAP selection is $\sim 89\%$.
The middle panel shows the two-dimensional completeness map for HectoMAP red galaxies. 
The survey completeness is fairly homogeneous over the entire field. 
There are a few streaks of low completeness around the edges of the survey,
 but even in these regions the completeness exceeds $\sim70\%$. 
We show a similar two-dimensional completeness map for 
 SDSS Main Sample galaxies plus HectoMAP red galaxies with $r_{petro, 0} < 17.77$ in the lower panel. 
Only $\sim 10\%$ of the objects satisfy the HectoMAP selection. 
We note that the SDSS sample is less uniform than the HectoMAP sampling.
The SDSS spectroscopic survey is patchy near bright stars or the high-density regions due to the fiber collision. 

% ======================================
% Figure 2 - Figure \ref{zsample}
% ======================================
\begin{figure}
\centering
\includegraphics[scale=0.49]{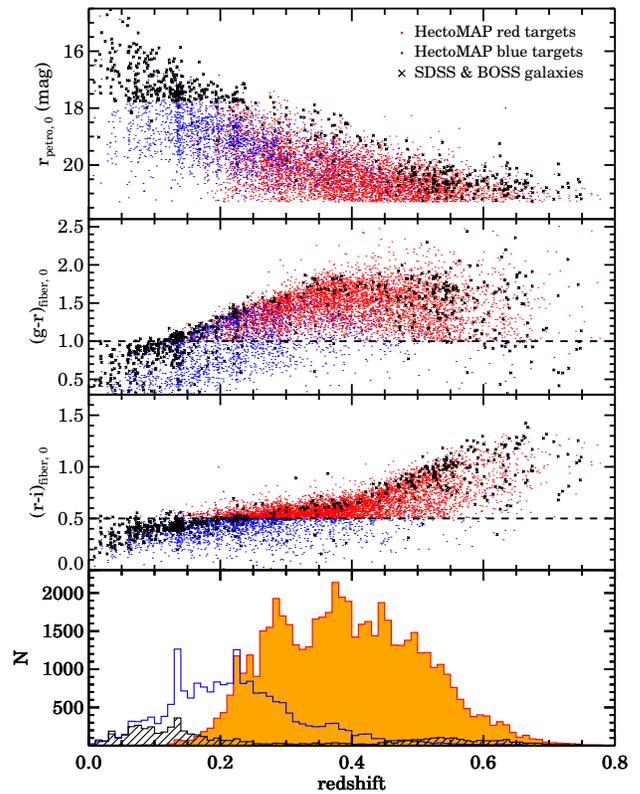}
\caption{
(Top) $r-$ band magnitudes as a function of redshift
 for HectoMAP red galaxies (red), 
 HectoMAP blue galaxies (blue), and 
 HectoMAP galaxies with SDSS spectra (black). 
(Middle) Same as the top panel, 
 but for  $(g-r)_{fiber, 0}$ and $(r-i)_{fiber, 0}$ colors. 
 The dashed lines display the HectoMAP color cuts. 
(Bottom) Redshift distribution of HectoMAP red galaxies (red filled), 
 the blue galaxies (blue open), and 
 SDSS galaxies (black hatched).  }
\label{zsample}
\end{figure}
% ======================================

Figure \ref{zsample} shows the color selection for HectoMAP. 
For comparison, 
 we plot SDSS galaxies and blue galaxies from HectoMAP.
The color selection we adopt effectively 
 selects galaxies with $z \gtrsim 0.2$.  

We use the red complete sample as the basis 
 for friends-of-friends (FoF) cluster identification at $z > 0.2$. 
At lower redshift, 
 the $r-i$ selection reduces the sampling density substantially 
 (Figure \ref{zsample}). 
Thus in this range we use the SDSS Main Sample 
 supplemented with the red HectoMAP galaxies.
 
After cross-identifying the FoF candidates with extended X-ray emission in the RASS
 we refine the cluster membership by applying the caustic technique \citep{Diaferio97, Diaferio99b, Serra13} 
 to all galaxies with redshifts in each cluster region. 
There are 26317 additional redshifts for objects bluer than the HectoMAP cuts. 
These objects include SDSS Main Galaxy sample, BOSS galaxies, and bluer galaxies 
 used to fill unused fibers in the Hectospec survey.  
This caustic analysis increases 
 the typical cluster membership by a factor of $\sim1.5$ compared to the FoF membership and
 enables more robust estimates of the cluster velocity dispersion and scale.

%=============================================================
\subsection{Search for X-ray emission in the ROSAT All-Sky Survey}\label{rosat}

For all the clusters found by the FoF algorithm in the HectoMAP survey,
 we searched for X-ray emission in the RASS \citep{Trumper93}.
To date, the RASS constitutes the only all-sky X-ray survey 
 conducted with an imaging X-ray telescope. 
The typical flux limit for a detection of at least $\sim 2.5 \sigma$ is 
 $\sim 3 - 5 \times 10^{-13}$ erg s$^{-1}$ cm$^{-2}$ in the 0.1 to 2.4 keV energy band. 
Detections of this low significance are only justified 
 because we search for emission at predefined positions. 
The source detection applied for the public RASS source catalog \citep{Voges99} 
 had a much higher significance threshold. 
The typical sky exposure in the RASS is of the order of 400 sec. 
For our purpose, 
 the survey is rather shallow, 
 and we expect a detection only for the most prominent systems.

In total, we searched for X-ray emission 
 at the position of 166 FoF clusters in the RASS allowing 
 in the first pass a coincidence radius up to 7 arcmin. 
We found 15 systems which showed significant X-ray emission.
These systems have more than 30 FoF members. 
The source detection and characterization follows 
 the techniques used for the construction of 
 the REFLEX II and NORAS II cluster surveys
 \citep{Bohringer13, Bohringer17, Chon12}. 
In the NORAS and REFLEX surveys 
 we ran our detailed source analysis at positions of a low threshold RASS X-ray source catalog \citep{Voges99}. 
Here we apply the same technique at the sky positions of the HectoMAP clusters. 
It uses the ROSAT PSPC detector energy channels 52 to 201 
 which roughly correspond to the energy band 0.5 to 2 keV, 
 because in this energy band the signal-to-noise ratio is maximized. 
At lower energies a large part of the X-ray emission is absorbed 
 by the interstellar medium of our Galaxy and the galactic X-ray background is high. 
Above 2 keV, 
 there is a sharp cut-off in the reflectivity of the ROSAT mirror.

To measure the X-ray fluxes and to characterize the X-ray sources,
 we apply the growth curve analysis method described in \citet{Bohringer00}.
In brief, the background subtracted cumulative source count rate 
 is determined with an increasing aperture, 
 and the fiducial source radius is identified with the location 
 where the count rate reaches a stable plateau. 
The X-ray source position is determined from
 the mean sky position of the source photons detected in an aperture of three arcmin radius. 
The offsets of the X-ray positions from the HectoMAP cluster positions are in Table \ref{xcl}.
 
In some cases, 
 nearby X-ray sources have to be deblended interactively to determine
 a proper source count rate
 \footnote{Deblending of a contaminated cluster X-ray source 
 can generally be performed for the following cases: 
 (i) the sources are clearly separable and the cluster source 
  can be well identified from the coincidence with the concentration of galaxies in the optical,
 (ii) the non-cluster source is clearly identified by a local contamination with 
  a different spectral hardness ratio (see \citealp{Chon12}), 
 (iii) the contaminating source can be clearly identified as 
 a point source superposed on the extended emission from the cluster.}.
The conversion from count rate to flux
 is obtained by folding X-ray spectra of hot thermal plasma through
 the instrument response function of the ROSAT instruments
 \footnote{The calculations are performed using the plasma spectral code
 xspec available from NASA HEASARC at https://heasarc.gsfc.nasa.gov/xanadu/xspec}. 
The parameters used for the spectrum calculations are the plasma temperature 
 obtained from the X-ray luminosity-temperature relation \citep{Bohringer12}
 a metal abundance of 0.3 solar, the known cluster redshift, 
 and the interstellar hydrogen column density taken from the 21cm sky maps of \citet{Dickey90}. 
The luminosities are determined for the rest frame 0.1 to 2.4 keV energy band.

In addition to the flux and the luminosity, 
 we determine two further quantities for each source: 
 the probability that the source is extended beyond the point spread function of the ROSAT instruments
 and the spectral hardness ratio (e.g. \citealp{Bohringer13}). 
The probability of having extended source emission is evaluated by means of a Kolmogorov-Smirnov test. 
A probability threshold of less than 1\% consistency with a point source is taken 
 as a significant detection of source extent. 
The hardness ratio of the sources is determined by means of the formula $ HR = (H-S) / (H+S)$, 
 where $H$ is the flux in the $0.5 - 2$ keV band and 
 $S$ the flux at $0.1 - 0.4$ keV. 
The observed hardness ratio is compared to the one 
 expected for thermal emission from a hot intracluster plasma
 enabling the identification of X-ray emission of possible other 
 contaminating X-ray sources contributing to the cluster X-ray flux. 
These extra source characterization parameters are 
 a great help in strengthening 
 the interpretation of the observed X-ray emission 
 as originating in the cluster's intracluster medium. 

%=============================================================
\section{CLUSTER IDENTIFICATION}\label{fof}

\subsection{The FoF algorithm}

We apply a FoF algorithm 
 to identify candidate galaxy clusters in HectoMAP. 
The FoF algorithm recursively links galaxies within 
 given linking lengths and bundles them 
 into candidate galaxy systems. 
The FoF algorithm is widely applied 
 because it identifies a unique set of group and cluster candidates in a sample 
 regardless of the geometry of clusters \citep{Berlind06}. 
Based on the FoF algorithm, 
 many group and cluster catalogs have been constructed 
 from a wide variety of spectroscopic surveys 
 \citep{Huchra82, Geller83, Ramella97, Ramella99, Barton96, Berlind06, Tago10, Robotham11, Tempel12, Tempel14, Tempel16, Sohn16}. 
 
We adopt a standard FoF algorithm 
 that connects neighboring galaxies with separate spatial and radial velocity linking lengths 
 \citep{Huchra82, Geller83}.
This application identifies friends of a galaxy 
 if the transverse ($\Delta D_{ij}$) and radial ($\Delta V_{ij}$) separations 
 are smaller than the selected fiducial linking lengths. 
Here, the separations between two galaxies ($i$ and $j$) are 
\begin{equation} 
\Delta D_{ij} =  {\rm tan} (\theta_{ij}) (D_{c, i} + D_{c, j}) / 2,
\end{equation}
 and
\begin{equation} 
\Delta V_{ij} = |D_{c, i} - D_{c, j}|, 
\end{equation}
 where $\theta_{ij}$ is an angular separation between two galaxies and
 $D_{c}$ is the comoving distance at the redshift of galaxy. 
 
The choice of linking length is a critical issue.  
If the linking lengths are too tight, 
 only compact systems are identified and 
 looser galaxy systems are broken into smaller fragments.
In contrast, 
 the algorithm links physically distinct systems or 
 even unrelated segments of the large-scale structure into a single cluster
 if the linking lengths are too generous.
\citet{Diaferio99a} also show that 
 properties of the candidate clusters 
 including the number of members, size, and velocity dispersion vary
 depending on the linking length. 
 
Previous studies that identify clusters based on a FoF algorithm 
 often choose linking lengths 
 related to the number density of galaxies ($\overline{n}_{g}$) at each redshift. 
In other words, the FoF identifies local overdensities as candidate systems.
Here we follow this approach for HectoMAP.

% ======================================
% Figure 3 - Figure \ref{dmean}
% ======================================
\begin{figure}
\centering
\includegraphics[scale=0.49]{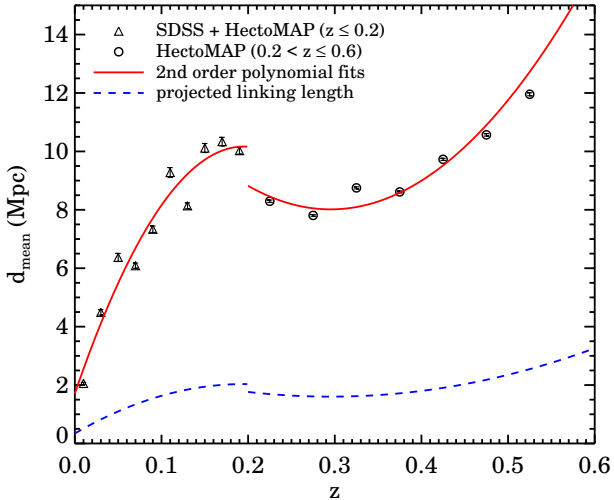}
\caption{Mean separation ($d_{mean}$) of the spectroscopically sampled red galaxies in HectoMAP (black circles). 
The red solid line shows the 2nd order polynomial fit for $0.2 < z \leq 0.6$.
The blue dashed line is the projected linking length ($0.2 \times d_{mean}$)
 we apply for identifying clusters in the higher redshift subsample
 as a function of redshift. 
The radial linking length is the same as $d_{mean}$. }
\label{dmean}
\end{figure}
% ======================================  

To apply the FoF algorithm to HectoMAP,
 we first compute the mean separation of galaxies 
 in the complete red-selected sample as a function of redshift.
Figure \ref{dmean} shows the mean separation ($\overline{n}_{g}^{-1/3}$) 
 as a function of redshift.
For galaxies at $z \leq 0.2$, 
 we count the number of the SDSS galaxies and the HectoMAP red galaxies 
 in redshift bins $\Delta z = 0.02$ and divide by the comoving volume. 
In this redshift range, 
 most redshifts of galaxies in the HectoMAP region are from SDSS.  
For galaxies at $0.2 < z \leq 0.6$, 
 we calculate the mean separation of the HectoMAP red galaxies in redshift bins $\Delta z = 0.05$. 
 
At both $z \leq 0.2$ and $0.2 < z \leq 0.6$, 
 the mean separation of galaxies increases smoothly as a function of redshift. 
The solid red curves show 
 2nd order polynomial fits to the mean separations. 
We then apply variable linking lengths
 according to these fits to the 
 effective survey number densities for each redshift range. 
  
The variable linking lengths related to the survey number density are 
\begin{equation}
 \Delta D  = b_{\bot} \overline{n}_{g}^{-1/3},
\end{equation}
and
\begin{equation}
|\Delta V| = b_{\|} \overline{n}_{g}^{-1/3},
\end{equation}
 where $b_{\bot}$ and $b_{\|}$ are scaling factor for the transverse and radial linking lengths. 
This application identifies clusters as over-densities at different redshifts. 
The $b_{\bot}$ determines the system over-density according to
\begin{equation}
\frac{\delta n}{n} = \frac{3}{4 \pi b_{\bot}^3} - 1
\end{equation}
\citep{Huchra82, Geller83, Diaferio99a, Duarte14}. 

Although many studies have applied the FoF algorithm with variable linking lengths, 
 there is no strong agreement on the choice of linking lengths \citep{Duarte14}. 
Previous studies employ $b_{\bot}$ ranging from 0.06 to 0.23 and 
 a fixed $b_{\bot}/b_{\|}$ ratio $\sim5 - 10$. 
Several studies test the choice of linking lengths 
 by measuring the group completeness and reliability in
 comparison with mock group catalogs derived from N-body simulations
 \citep{Frederic95, Merchan02, Eke04, Berlind06, Robotham11}. 
However, 
 \citet{Berlind06} argue that no combination of $b_{\bot}$ and $b_{\|}$ 
 identifies clusters that simultaneously recover
 the halo multiplicity function and 
 the distribution of projected size and velocity dispersion for these systems. 

Here, we use linking lengths 
 $b_{\bot} = 0.2$ and $b_{\|} = 1.0$ ($b_{\bot} / b_{\|} = 5$). 
Our $b_{\bot}$ identifies systems 
 that have $\delta n/n \sim 29$. 
The limiting overdensity is fixed throughout the redshift range.
The set of linking lengths is somewhat generous compared to previous studies.  

We use the FoF algorithm only
 as a basis for an X-ray detected cluster catalog in the HectoMAP region.
We would rather include false candidates than exclude real ones.
We next show that this approach recovers the relevant set of X-ray clusters
 in a separate densely surveyed region with essentially full X-ray coverage \citep{Starikova14}. 

We test the FoF on the Smithsonian Hectospec Lensing Survey (SHELS) survey \citep{Geller10, Geller12, Geller14, Geller16}, 
 a dense complete redshift survey with no color selection and 
 covering two well-separated 4 deg$^{2}$ fields (F1 and F2) of the Deep Lens Survey (DLS, \citealp{Wittman02}). 
We use the F2 survey alone for this investigation 
 because it has essentially complete X-ray coverage of all of the possible massive systems 
 in the appropriate redshift range.
The F2 survey is $\sim95\%$ complete to R $\leq 20.6$ with a number density of $\sim 3200$ galaxies deg$^{-2}$.

In total, there are 26 X-ray extended sources detected
 based on {\it Chandra} and XMM X-ray data \citep{Starikova14};
 18 of these X-ray sources are in the redshift range $0.2 < z \leq 0.6$ of interest 
 for testing our approach to HectoMAP X-ray clusters. 
Here, we select 12 X-ray sources within $0.2 < z \leq 0.5$ 
 for testing our cluster identification procedure
 because the F2 magnitude limit $R \leq 20.6$ is $\sim0.4$ magnitudes brighter than the HectoMAP limit. 
The brighter F2 limit precludes FoF cluster detection at $z \gtrsim 0.5$.
The X-ray flux limit for the 12 X-ray extended sources with $z < 0.5$ 
 is $4.3 \times 10^{-15}$ erg cm$^{-2}$ s$^{-1}$
 within $0.5 - 2.0$ keV energy band. 
This X-ray flux limit is nearly two orders of magnitude fainter 
 than the one we reach for the HectoMAP systems. 

\citet{Starikova14} show that ten of these sources have 
 optical counterparts identified as clusters 
 in the SHELS spectroscopic survey \citep{Geller10} or in the DLS weak lensing analysis \citep{Wittman06}. 
Here, we examine the number of F2 X-ray extended sources 
 recovered by the FoF algorithm with the HectoMAP choice of linking lengths. 

% ======================================
% Figure 4 - Figure \ref{f2cone}
% ======================================
\begin{figure}
\centering
\includegraphics[scale=0.43]{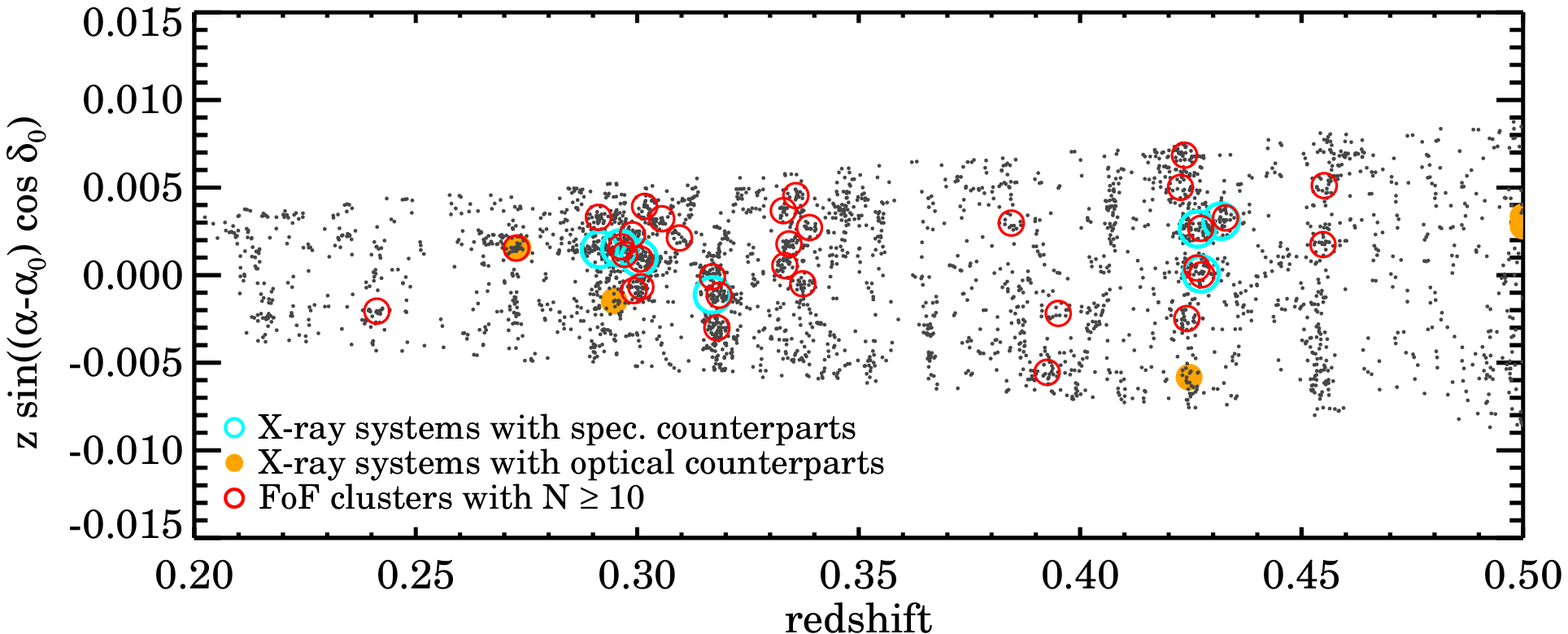}
\caption{
Cone diagram for the SHELS F2 survey subsample 
 with $R \leq 20.6$, $(g-r)_{fiber,0} > 1.0$, $(r-i)_{fiber, 0} > 0.5$
 projected in the R.A. direction. 
Cyan and yellow circles indicate
 extended X-ray sources in F2
 with spectroscopic counterparts \citep{Geller10} and 
 photometric counterparts \citep{Wittman06, Starikova14}, respectively. 
Red circles display FoF cluster candidates containing more than 10 members. 
Every extended X-ray source with a spectroscopic counterpart is recovered by the FoF algorithm.}
\label{f2cone}
\end{figure}
% ======================================  

Figure \ref{f2cone} shows the cone diagram of 
 the subset of F2 galaxies selected according to the HectoMAP red galaxy prescription
 within $0.2 < z \leq 0.5$.
In Figure \ref{f2cone}, 
 the cyan circles display the locations of seven X-ray extended sources 
 with spectroscopically identified counterparts from \citet{Geller10}. 
We also show the X-ray extended sources (yellow circles in Figure \ref{f2cone})
 matched with weak lensing peaks \citep{Wittman06}
 or with optical concentrations on the sky in the DLS images \citep{Starikova14}. 
The redshifts of these X-ray sources with photometrically identified counterparts
 may not be accurate because they are based on photometric redshifts. 
  
In summary, 
 the FoF algorithm identifies six systems with $N \geq 10$ 
 that match extended X-ray sources in the redshift range $0.2 < z \leq 0.5$. 
All X-ray sources with spectroscopically identified counterparts have FoF counterparts; 
 one large FoF system at $z = 0.29$ includes two of these X-ray extended sources. 
\citep{Starikova14} show that with detailed modelling the FoF system resolves 
 in a way that corresponds to the two extended X-ray sources.
These massive systems in A781 complex ($z=0.298$) are very close together on the sky and
 they are difficult to separate in the spectroscopic data 
 without a more detailed analysis \citep{Geller10}. 
Among the four X-ray extended sources lacking FoF counterparts, 
 all are identified only as apparent photometrically identified overdensities in \citet{Starikova14}.
One of these sources does match an FoF system with $N = 5$. 
We find no FoF systems associated with the other three X-ray sources 
 either because their redshifts are incorrect or because they are contaminated by unresolved X-ray point sources. 
The performance of the FoF approach 
 over the range accessed by the redshift survey supports 
 application of this approach to detection of X-ray clusters 
 to a brighter X-ray flux limit in the HectoMAP data.

To identify cluster candidates over the full redshift range of HectoMAP, 
 we apply the FoF algorithm to separate samples 
 for $z \leq 0.2$ and for $0.2 < z \leq 0.6$. 
For $z < 0.2$, 
 we select the combined SDSS Main Sample galaxies and HectoMAP red galaxies 
 as the basic galaxy sample. 
For $0.2 < z \leq 0.6$, 
 we only use the HectoMAP red galaxies to identify cluster candidates.
We follow this approach 
 because the HectoMAP $r-i$ cut removes most galaxies at $z \lesssim 0.2$ 
 thus compromising cluster identification 
 in this range based on HectoMAP galaxies alone.
  
% ======================================
% Figure 5 - Figure \ref{zhist}
% ======================================
\begin{figure}
\centering
\includegraphics[scale=0.49]{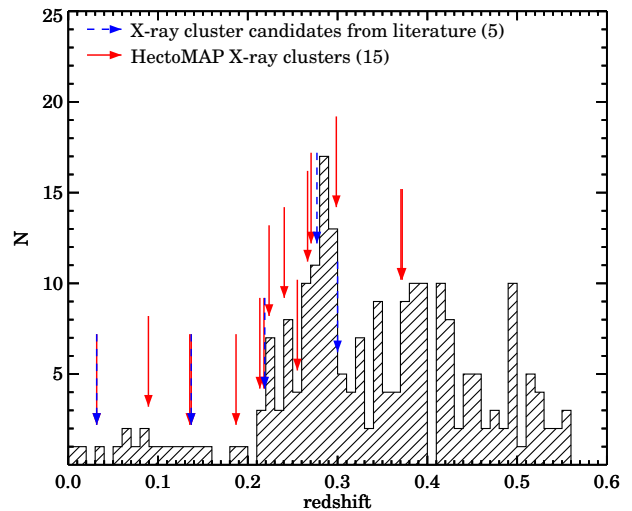}
\caption{
Redshift distribution of FoF cluster candidates in HectoMAP. 
Red and blue arrows mark the redshifts of 
 the HectoMAP X-ray clusters and clusters from the literature, respectively.}
\label{zhist}
\end{figure}
% ======================================  
  
The FoF algorithm detects 166 candidate systems each consisting of at least 10 galaxies. 
Figure \ref{zhist} shows the redshift distribution of these candidate systems. 
In Figure \ref{zhist}, 
 we mark the 15 clusters with associated extended X-ray emission. 
The approximate flux limit for X-ray detected systems is $\sim 3 \times 10^{-13}$ erg cm$^{-2}$ s$^{-1}$. 
None of the HectoMAP clusters at $z > 0.4$ are detected in the RASS. 
Detection of them would require high X-ray luminosity (cluster mass).  
For example, 
 the fluxes of the three $z>0.4$ X-ray clusters in F2 field are $1.6, 1.9, 0.2 \times 10^{-13}$ erg s$^{-1}$ cm$^{-2}$, 
 below our X-ray detection limit. 

\subsection{X-ray Counterpart of FoF clusters}\label{Xray}

Of the clusters detected with the FoF algorithm 15 systems show 
 significant X-ray emission in the RASS. 
The number of net source photons found are 
 in the range from 6 to 57 
 (with the exception of HMxcl141341.4+433925 with 128 source photons, which is discussed below). 
The RASS sky has a low background and 
 the detections are significant even with low number of photons, 
 but the characterization of the sources becomes difficult with few photons. 
Thus we are not expecting to find all the cluster sources 
 to have significantly extended X-ray emission. 
Indeed, five of the clusters do not fulfill our criterion for source extent. 
These clusters are all detected with less than 30 photons. 
Therefore, the failure to establish a source extent is not 
 an argument against their cluster nature. 
All the sources with more than 30 photons show a clear extent.

All clusters show 
 a spectral hardness ratio consistent with that expected 
 from thermal emission of a hot intracluster plasma. 
The only exception is the cluster HMxcl141341.6+433925, 
 which shows spectral parameters indicating that 
 the X-ray emission is too soft for intracluster plasma emission. 
There is no signature of a positional difference of a softer and harder X-ray source. 
Therefore, the most likely explanation for the softness of the X-ray source is that 
 there is the contribution of an AGN in the brightest cluster galaxy 
 located exactly at the X-ray center and X-ray maximum of the X-ray source. 
Since the X-ray source is clearly extended, 
 not all the emission can come from the AGN and 
 we can put an upper limit that no more than 50\% of the X-ray flux can come from the AGN.
Fitting a point source to the source profile of HMxcl141341.6+433925 
 we obtain a maximum flux for the point source of $0.27 \times 10^{-12}$ erg s$^{-1}$ cm$^{-2}$,
 $\sim 49$\% of the total source flux.

\subsection{Caustic Method and Cluster Membership}\label{caustic}

% ======================================
% Figure 6 - Figure \ref{rvxcl}
% ======================================
\begin{figure*}
\centering
\includegraphics[scale=0.6]{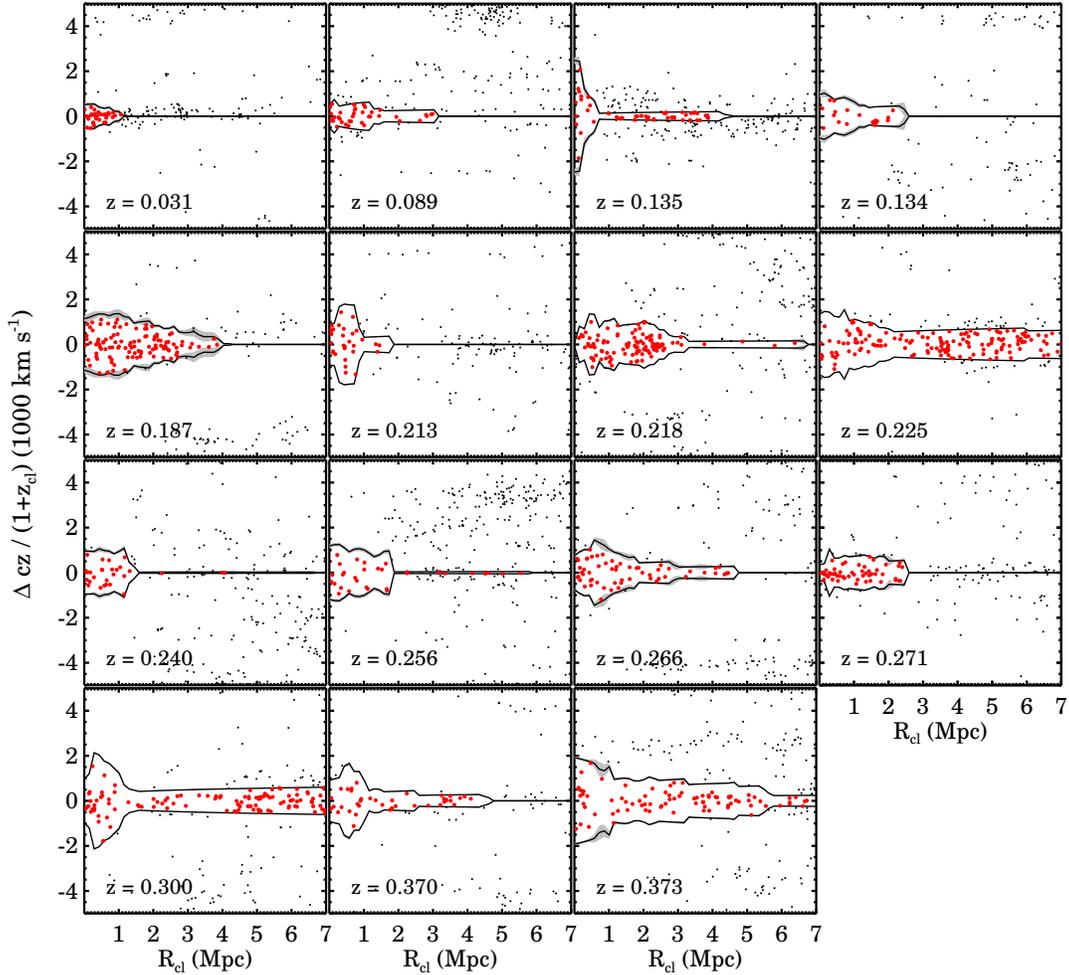}
\caption{
Rest-frame clustercentric radial velocities
 vs. projected clustercentric distance (R-v diagrams) 
 for HectoMAP X-ray clusters sorted by redshifts. 
Black points are galaxies along the line-of-sight and 
 red points are clusters members within the caustics. 
The black solid lines show the caustics 
 and the gray shaded regions indicate the uncertainty in the caustic estimate. }
\label{rvxcl}
\end{figure*}
% ======================================

We use the caustic technique to refine 
 the definition of the X-ray clusters
 and to determine the cluster membership. 
We run the FoF algorithm on the complete red galaxy sample alone. 
However, the Hectospec survey contains many objects 
 outside the color cuts 
 that are useful for calculating the properties of the clusters. 
We incorporate these objects in the caustic analysis.
  
The caustic technique is a powerful tool for identifying cluster members 
 \citep{Diaferio97, Diaferio99b, Serra13}.
This non-parametric technique determines the boundaries of clusters (caustics)
 which delimit the location of the cluster members. 
\citet{Serra13} test the membership determination based on the caustic technique 
 with mock catalogs containing $\sim1000$ galaxies 
 within a field of view of 12 Mpc/h on a side at the cluster location. 
Their simulations show that $\sim 92\%$ of cluster members are recovered for clusters containing 
 at least $\sim 50$ members within $R_{200}$;
 the contamination from interlopers is $\sim 3\%$. 
Because the X-ray detected HectoMAP clusters have a few tens of members, 
 we expect the caustic technique to identify cluster members 
 with a similar success rates. 

The caustic technique is also effective 
 for disentangling structures along the line-of-sight; 
 these structures may be falsely linked by the generous radial linking length we use. 
For example,
 \citet{Rines13} distinguish the pair of clusters
 A750 and MS0906+11, based on the caustic technique.
The radial separation between two clusters is $\sim3250~\kms$. 
The caustics for the two clusters clearly segregate members of the superimposed clusters
 (Figure 9 of \citealp{Rines13}). 

We calculate caustics based on galaxies with redshifts
 within the 30 arcmins of the cluster centers determined by the FoF algorithm. 
In determining cluster boundaries, 
 the caustic technique also revises the cluster center and 
 the mean redshift. 
Hereafter, we use the cluster centers and redshifts 
 from the caustic technique. 

Figure \ref{rvxcl} displays 
 the rest-frame clustercentric velocity
 as a function of projected clustercentric distance, 
 the R-v diagram, for the galaxies in the 15 HectoMAP FoF clusters with X-ray counterparts. 
The R-v diagrams show clear concentrations around 
 the center of each cluster.
The solid lines in each panel 
 mark the boundaries of the clusters identified by the caustic technique. 
The galaxies within these caustic patterns are cluster members. 

The caustic technique also provides the cluster mass profile
 \citep{Diaferio97, Diaferio99b, Serra11}. 
Based on this mass profile, 
 we compute the characteristic cluster mass and size, 
 i.e., $M_{200}$ and $R_{200}$. 
The mass of HectoMAP X-ray clusters ranges from 
 $2 \times 10^{13}$ M$_{\odot}$ to $4 \times 10^{14}$ M$_{\odot}$
 comparable with the CIRS \citep{Rines06} and HeCS \citep{Rines13} samples. 
We estimate the velocity dispersion of each cluster
 using the method given in \citet{Danese80}. 
Hereafter, $\sigma_{cl}$ denotes the rest frame line-of-sight velocity dispersion 
 for cluster members within $R_{200}$.

%=============================================================

\section{A Catalog of HectoMAP FoF clusters with X-ray Emission}\label{cat}

% ======================================
% Figure 7 - Figure \ref{hsc}
% ======================================
\begin{figure*}
\centering
\includegraphics[scale=0.5]{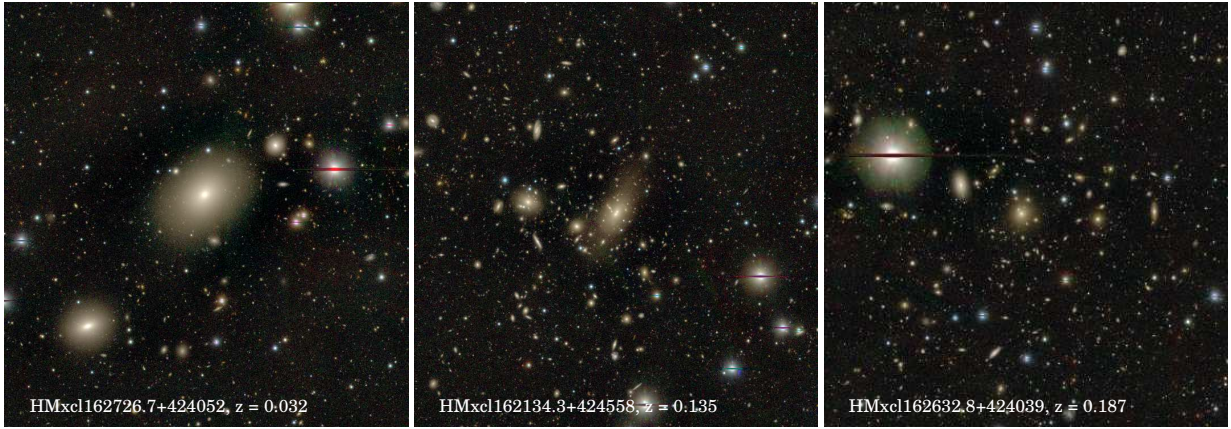}
\caption{
{\it Subaru}/Hyper Supreme-Cam images of three HectoMAP X-ray clusters centered on the BCGs of the clusters.
The image sizes are 4 arcmin by 4 arcmin. 
The X-ray centers are near the BCGs. } 
\label{hsc}
\end{figure*}
% ====================================== 

We identify 15 robust clusters in HectoMAP 
 based on the FoF method combined with the identification of X-ray emission in the RASS. 
Hereafter, we refer these HectoMAP FoF clusters with X-ray counterparts 
 as HectoMAP X-ray clusters.
We identify the brightest cluster galaxies (BCGs) among 
 the spectroscopically determined cluster members. 
Six out of 15 systems contain a galaxy brighter than the BCGs within $R_{cl} < 3\arcmin$ and without a redshift. 
These objects are most likely foreground blue objects excluded by the HectoMAP selection.
The SDSS photometric redshift estimate for these objects suggests that they are all foreground objects. 
The projected distances between these galaxies and the cluster center is also larger 
 than normal for a BCG.
The offset between X-ray emission for the systems and the BCGs of the clusters 
 is consistent with the ROSAT PSF ($\sim 2\arcmin$, \citealp{Boese00}).
For one system, HMxcl145913.1+425808, 
 the offset between the X-ray emission and the BCG is slightly larger ($\sim 2.24\arcmin$) than the PSF, 
 but the offset from the cluster center determined from the caustics ($\sim 0.96\arcmin$) is still within the PSF.  

Figure \ref{hsc} displays 
 examples of three HectoMAP X-ray clusters that lie within 
 the publicly released {\it Subaru}/Hyper Suprime-Cam (HSC) imaging footprint.
The HSC images clearly demonstrate that 
 the HectoMAP X-ray clusters are generally massive systems 
 containing BCGs surrounded by plenty of members. 
Even in SDSS images, 
 it is evident that the other HectoMAP X-ray clusters are also massive systems.

Table \ref{xcl} lists the properties of the HectoMAP X-ray clusters
 sorted by redshift. 
The table contains the cluster ID, the position of the cluster center,
 the cluster mean redshift, the number of members within the caustics,
 the center of the extended X-ray emission, 
 the offset between the BCG and the center of the X-ray emission, and 
 the X-ray luminosity. 
The X-ray luminosity is given in the 0.1 to 2.4 keV energy band in the cluster rest frame. 
Table \ref{xcl_dyn} summarizes 
 the dynamical properties of the HectoMAP X-ray clusters 
 including $\sigma_{cl}$, $R_{200}$, and $M_{200}$. 
We also list the 1036 cluster members with their redshifts and the redshift source in Table \ref{xclmem}. 
 
% ======================================
% Figure 8 - Figure \ref{cmd}
% ======================================
\begin{figure*}
\centering
\includegraphics[scale=0.6]{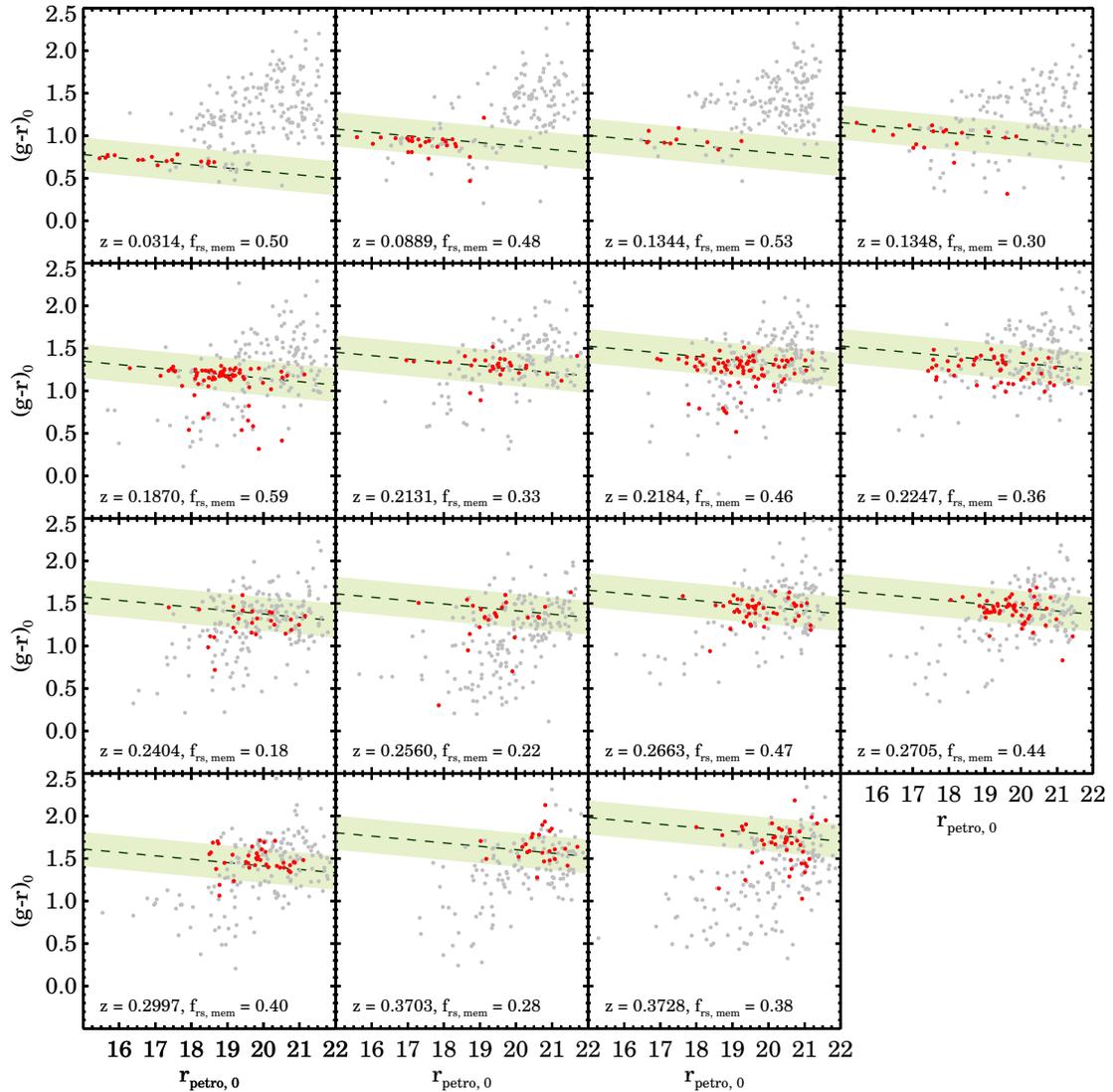}
\caption{
$g-r$ vs. $r$ color-magnitude diagram of 
 the HectoMAP X-ray clusters sorted by redshifts. 
Red and gray circles show cluster members and spectroscopic targets
 within $10\arcmin$, respectively. 
Shaded regions mark the red-sequence (dashed line) $\pm 0.1$. 
$f_{rs, mem}$ shows the member fraction with respect to the 
 number of spectroscopic targets in the red-sequence. }
\label{cmd}
\end{figure*}
% ======================================  
 
Clusters of galaxies are often identified by the red sequence 
 (e.g. \citealp{Gladders00, Rykoff14, Oguri17}). 
However, not all galaxies on the red sequence are cluster members \citep{Rines08, Sohn17}.  
We examine the color magnitude diagram for 
 the HectoMAP cluster regions to explore this issue.  
Figure \ref{cmd} shows the observed $g-r$ color versus $r-$band magnitude
 of the spectroscopically sampled galaxies 
 within 10 arcmins of each HectoMAP X-ray cluster.
We identify the red-sequence
 by assuming a slope of $-0.04$ in color-magnitude space 
 following \citet{Rines13}. 
We classify objects within $\pm 0.1$ of the relation 
 as red-sequence members. 
Among the cluster members identified by the caustic technique, 
 the fraction on the red-sequence ranges from 55\% to 92\%
 consistent with previous spectroscopic surveys of massive clusters overlapping this redshift range.

% ======================================
% Figure 9 - Figure \ref{cone}
% ======================================
\begin{figure*}
\centering
\includegraphics[scale=0.4]{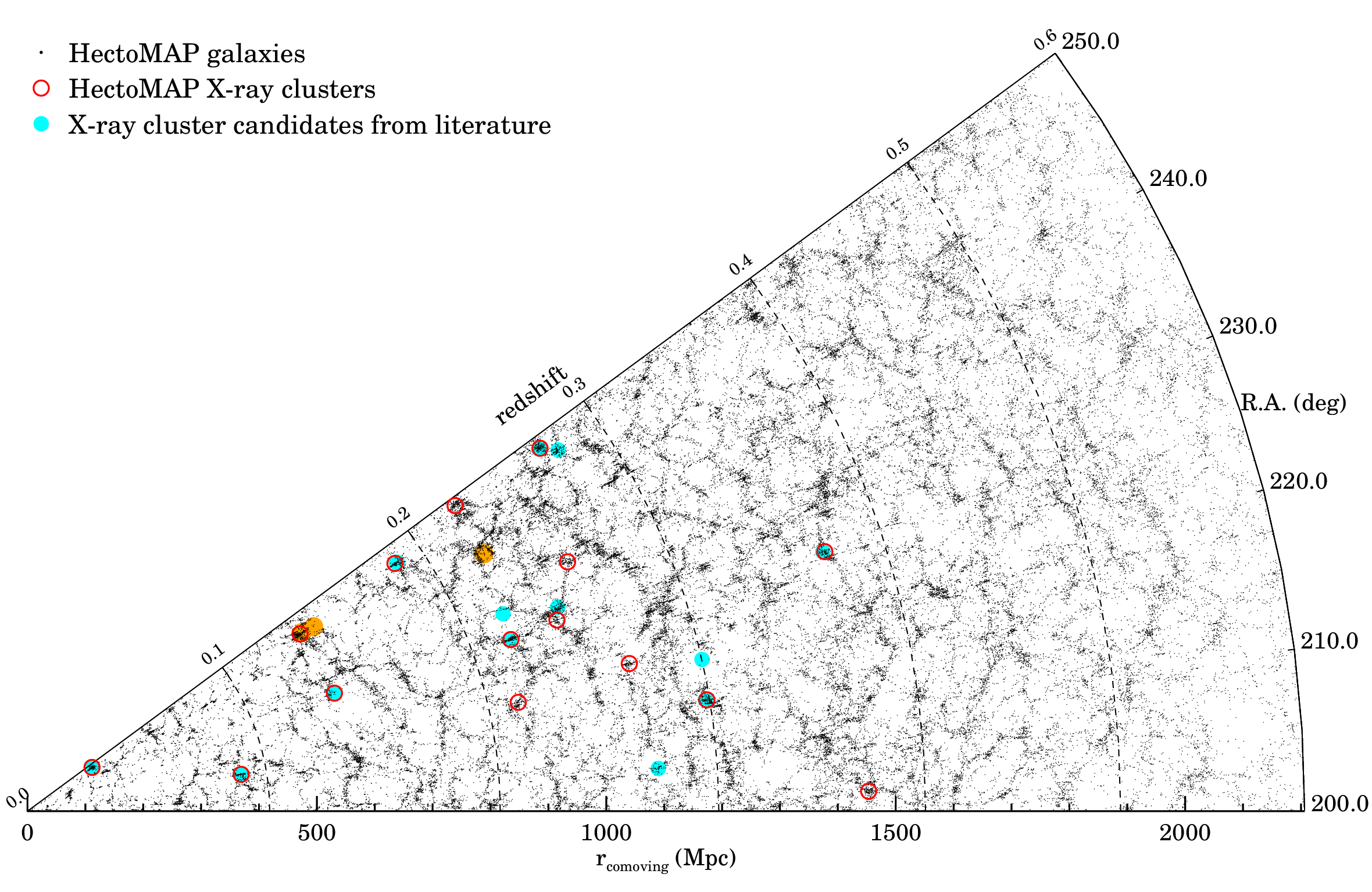}
\caption{Cone diagram for the HectoMAP region. 
Black dots indicate the HectoMAP red galaxies. 
Red circles show HectoMAP X-ray clusters.
Blue circles mark X-ray cluster candidates from the literature. }
\label{cone}
\end{figure*} 
% ====================================== 

Among the galaxies projected onto the red sequence in the cluster field, 
 the fraction of HectoMAP X-ray cluster members ($f_{rs, mem}$) is remarkably low: 
 17\% to 54\% with a median of 36\%.
The quantity $f_{rs, mem}$ is the ratio 
 between the number of spectroscopically identified cluster members and 
 the number of spectroscopic targets on the red sequence. 
We compute the $f_{rs, mem}$ using all cluster members regardless of their apparent magnitude, 
  rather than limited to $r \leq 21.3$. 
The $f_{rs, mem}$ changes little when we estimate using the cluster members brighter than the HectoMAP magnitude limit ($r \leq 21.3$). 
The lower membership fraction simply reflects 
 the higher median redshift of the HectoMAP X-ray clusters 
 relative to previous spectroscopic samples 
 where this comparison has been made. 
Most of the objects that contaminate the red sequence are background. 

Figure \ref{cone} shows a cone diagram for the HectoMAP sample
 with $r_{\rm petro, 0} < 21.3$. 
The red circles on Figure \ref{cone} 
 show the location of the HectoMAP X-ray clusters. 
For comparison, 
 we also show the positions of previously reported X-ray clusters in the literature 
 (blue circles, see the details in Section \ref{lit}). 
The HectoMAP X-ray clusters are all embedded in dense structures. 
On the other hand, 
 many dense structures contain no HectoMAP X-ray clusters 
 mainly as a result of the lack of extended X-ray extended emission. 
In a forthcoming paper, 
 we will include the full FoF catalog for HectoMAP and will analyze it in detail (Sohn et al. in prep.). 
In general clusters in this catalog mark all of the densest regions in the survey.

%=============================================================
\section{DISCUSSION}\label{discussion}

Combining the dense redshift survey HectoMAP 
 with the RASS enables construction of 
 a robust catalog of X-ray clusters. 
These HectoMAP X-ray clusters contain $\sim 50$ members (median)
 within the redshift survey (Table \ref{xcl}).
The virtue of the cluster survey based on spectroscopic survey data
 is reduction of contamination by foreground and background structures. 
In particular, the caustic method efficiently eliminates 
 non-members along the line-of-sight. 

There are several cluster surveys covering the HectoMAP field 
 including red sequence detection (e.g. redMaPPer, \citealp{Rykoff14}), 
 identification of over-densities based on photometric redshifts \citep{Wen09}, and 
 identification of X-ray sources in partial overlapping surveys (reference in Table \ref{pxcl}). 
HectoMAP thus provides an opportunity  for studying 
 the spectroscopic properties of the clusters in the previous literature (Section 5.2). 
\citet{Sohn17b} is an extensive investigation of redMaPPer cluster candidates 
 based on the HectoMAP redshift survey. 
Here, we limit our discussion to X-ray detected systems (Section 5.1.)

We investigate the X-ray scaling relation 
 for the HectoMAP X-ray clusters and 
 compare it with relations from the literature in Section 5.2. 
Finally, we estimate the frequency of X-ray clusters 
 to the depth of the RASS in Section 5.3. 
The cluster masses and X-ray luminosities provide 
 a route to the estimated number density of X-ray clusters 
 over a larger region that has been possible before to this depth. 
This estimate is a useful guideline 
 for the next generation X-ray surveys (e.g. e-ROSITA).

\subsection{Previous X-ray cluster candidates in HectoMAP}\label{lit}

To investigate the spectroscopic properties of previously known X-ray clusters in the HectoMAP field, 
 we first search the literature. 
Several surveys detect X-ray clusters in HectoMAP
 \citep{Vikhlinin98, David99, Bohringer00, Lubin04, Burenin07, Horner08, Voevodkin10}. 
The MCXC catalog \citep{Piffaretti11} and the BAX catalog \footnote{http://bax.ast.obs-mip.fr}
 facilitate the search. 
Within the MCXC and the BAX catalogs, 
 we find five and eight X-ray cluster candidates in the HectoMAP field, respectively. 
The clusters from the MCXC catalog are identified based on ROSAT 
 and those from the BAX catalog are from XMM or ASCA. 
Some of these clusters overlap leaving 
 a total of ten X-ray clusters from the MCXC and BAX catalogs. 
 
\citet{Wen09} also list galaxy cluster candidates with X-ray counterparts 
 in the HectoMAP field. 
Based on SDSS DR6 data, 
 they identify cluster candidates as over-densities  
 within a 0.5 Mpc radius and within the photometric redshift range 
 $|\Delta (z_{phot} - z_{BCG})| < 0.04~(1 + z_{BCG})$. 
They match their photometrically identified cluster candidates 
 with the ROSAT point source catalog and 
 provide a list of cluster candidates with X-ray point source counterparts
 (their Table 2).
They identify eight X-ray cluster candidates in the HectoMAP field;
 three of them overlap the systems from the MCXC and BAX catalogs. 

Table \ref{pxcl} lists the 15 X-ray cluster candidates in the HectoMAP field 
 from the literature. 
The positions and redshift of clusters are from the literature. 
The X-ray flux and luminosity are based on the ROSAT band (0.1 - 2.4 keV). 
For those objects with X-ray photometry in other bands (e.g. 0.5 - 2.0 keV), 
 we converted to the ROSAT band using the PIMMS 
 \footnote{https://heasarc.gsfc.nasa.gov/docs/software/tools/pimms.html}. 
Among 15 cluster candidates from the literature, 
 eight systems match HectoMAP X-ray clusters. 

For the remaining 7 cluster candidates, 
 we revisit the previously known X-ray sources 
 that are not associated with HectoMAP X-ray clusters in the RASS data.
The RASS yields only upper limits on the X-ray fluxes for four systems (Table \ref{pxcl}). 
We do not detect any X-ray flux for three sources, 
 MCXC1515.6+4350 \citep{Vikhlinin98}, 
 GHO1602+4312 \citep{Lubin04} and 
 MCXC1429.0+4241 \citep{Horner08}.
 
The previous detections of MCXC1515.5+4346 and MCXC1515.6+4350 are confusing. 
These sources originate from \citet{Vikhlinin98} who identify two X-ray sources:
 VMF 168 (R.A., Decl., z = 15:15:32.5, +43:46:39, $\sim0.26$) and 
 VMF 169 (15:15:36.8, +43:50:50. $\sim0.14$). 
Later, \citet{Burenin07} and \citet{Voevodkin10} list only one X-ray source (15:15:33.0, +43:46:35) 
 near VMF 168, but at $z = 0.137$, similar to VMF 169. 
Indeed, we detect one system (HMxcl151550.0+434556) at $z = 0.137$;
 this cluster matches VMF 168 and the X-ray source from \citet{Burenin07}. 
We suspect that the redshift of VMF 168 listed in \citet{Vikhlinin98} is incorrect.

We do not identify X-ray emission around VMF 169 in the RASS data.
We find an overdensity of galaxies at $z \sim 0.243$ near VMF 169, 
 but the center of the overdensity is significantly offset ($\sim 11\arcmin$) from the published location of VMF 169.
In conclusion, there is only one significant extended X-ray source at $z = 0.137$,
 consistent with the source from \citet{Burenin07} and \citet{Voevodkin10}. 
 
% ======================================
% Figure 10 - Figure \ref{rvpxcl}
% ======================================
\begin{figure}
\centering
\includegraphics[scale=0.38]{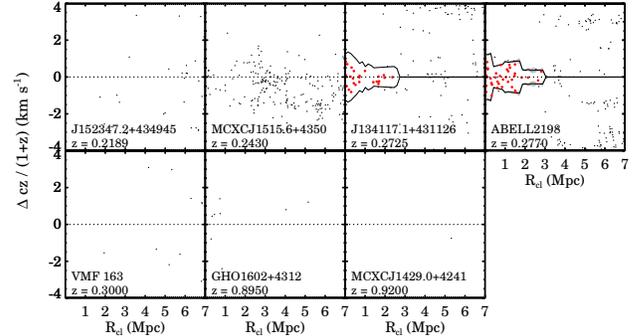}
\caption{
R-v diagrams for 7 X-ray cluster candidates in the HectoMAP region
 listed in the literature.
Cluster candidates here are not included among the HectoMAP X-ray clusters.  
Red and black circles show cluster members and spectroscopic targets, respectively. 
The plots are centered on the known (photometric) redshifts of the cluster candidates. 
There is no spectroscopic evidence of cluster 
 for three systems: WHL J152347.2+434045, MCXCJ1515.6+4350 (VMF 169), and VMF 163. 
At $z > 0.6$, the redshift survey is sparse
 which limits spectroscopic detection of 
 GHO1602+4312 and MCXCJ1429.0+4241. }
\label{rvpxcl}
\end{figure}
% ====================================== 
 
Figure \ref{rvpxcl} shows R-v diagrams for seven previously identified X-ray cluster candidates
 that lack HectoMAP X-ray cluster counterparts. 
Because HectoMAP has few redshifts for galaxies with $z > 0.6$, 
 the R-v diagrams at these redshifts do not show structures associated with the reported X-ray sources. 
The R-v diagrams demonstrate that 
 the X-ray cluster candidates from the literature at lower redshift do not always have optical counterparts. 
We do not find obvious members associated with 
 WHL J152347.2+434945, MCXCJ1515.6+4350 and VMF 163. %J152348.2+434945. 
These objects could be matched with foreground or background X-ray sources like quasars. 
We can calculate caustics for two systems, 
 WHL J134117.1+431126 and Abell 2198. 
Although the FoF algorithm identifies the two systems, 
 the HectoMAP X-ray cluster catalog does not include these systems
 because we can measure only the upper limit of X-ray flux based on the RASS.  
We measure $\sigma_{cl}$ and $M_{200}$ for these systems (see caption of Table \ref{pxcl}). 

Abell 2198 is a mysterious case. 
The redshift of Abell 2198 is reported as $z = 0.0798$ \citep{Ciardullo83, Abell89}. 
\citet{David99} measured an upper limit on the X-ray flux from ROSAT 
 possibly coincident with this cluster. 
However, the R-v diagram shows no distinctive structure
 at the reported redshift even though SDSS covers this redshift range quite densely. 
Instead, we find a cluster at $z = 0.277$ (shown in Figure \ref{rvpxcl}). 
The brightest galaxy is offset from 
 the reported center of A2198 by $\sim2\arcmin$. 

We suspect that the previously published A2198 redshift was
 based on redshifts of a few foreground galaxies. 
Interestingly, \citet{Wen09} identify this cluster 
 based on photometric redshifts
 and report the cluster redshift as $z_{phot} = 0.284$, 
 remarkably close to the HectoMAP result.
\citet{Wen09} did not find the associated X-ray source 
 perhaps because of the positional offset.
Here, we examine the properties of A2198 
 including X-ray luminosity and velocity dispersion
 based on the HectoMAP redshift.  

The R-v diagrams in Figure \ref{rvxcl} and Figure \ref{rvpxcl}
 provide an estimate of the spectroscopic redshift for most of X-ray cluster candidates 
 from \citet{Wen09}.
The mean cluster redshift offset between the photometric and spectroscopic measures
 is $\Delta z_{{\rm phot} - {\rm spec}} = 0.015 \pm 0.008$ ($\sim 4500~\kms$). 
Considering the typical error in the photometric redshifts,
 the agreement is excellent.
 
Overall, the census of clusters in the literature 
 contains no additional systems above the RASS flux limit. 
This result suggests that 
 our catalog construction method yields a complete flux-limited sample. 
Furthermore, the R-v diagrams for the X-ray cluster candidates in the literature 
 underscore the importance of cross-checking cluster identification with dense spectroscopy. 
Some X-ray cluster candidates appear to be false detections, 
 but photometric redshifts do provide cleaner samples when the cluster candidates have an X-ray counterpart
 (e.g. \citealp{Wen09}).  

\subsection{Cluster Scaling Relation}\label{scl}

% ======================================
% Figure 11 - Figure \ref{msig}
% ======================================
\begin{figure}
\centering
\includegraphics[scale=0.49]{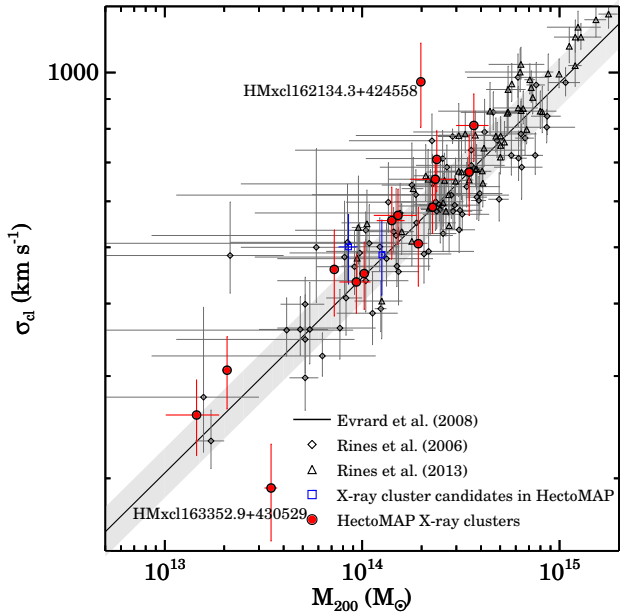}
\caption{
Velocity dispersion ($\sigma_{cl}$) vs. $M_{200}$ (dynamical mass within $R_{200}$)
 for the HectoMAP clusters. 
Red circles show the HectoMAP X-ray clusters and 
 blue circles display the previously identified X-ray clusters in HectoMAP. 
For comparison, 
 we show
 the CIRS clusters (diamonds, \citealp{Rines06}) and 
 the HeCS clusters (triangles, \citealp{Rines13}). 
The solid line shows the theoretical relation for dark matter halo 
 derived from cosmological simulations \citep{Evrard08}.
The gray shaded region indicates the standard deviation of the theoretical relation.}
\label{msig}
\end{figure}
% ====================================== 

Figure \ref{msig} displays the $M_{200} - \sigma_{cl}$ relation 
 for the HectoMAP X-ray clusters. 
We also show the relation for the X-ray cluster candidates from the literature.  
The HectoMAP clusters lie on the trend defined by 
 the larger, lower redshift CIRS \citep{Rines06} and HeCS \citep{Rines13, Rines16} samples.
 
We find two outliers: 
 HMxcl162134.3+424558 with $\sigma_{cl} \sim 963~\kms$ and 
 HMxcl163352.9+430529 with $\sigma_{cl} \sim 193~\kms$. 
We suspect that poor sampling of the HMxcl162134.3+424558 central region
 precludes measuring a reasonable velocity dispersion. 
The second cluster, HMxcl163352.9+430529, is puzzling. 
The R-v diagram looks reasonable and the $\sigma_{cl}$ based on 54 members should be robust. 
The low velocity dispersion of this system may be  
 due to poor sampling of the central region or an anisotropy of this system, 
 (i.e. we observe this cluster along its minor axis). 
A denser redshift might provide 
 a better understanding of the low $\sigma_{cl}$ for this cluster. 
 
Because the $M_{200}$ of a cluster is correlated with 
 the velocity dispersion of cluster members,
 the tight correlation between $M_{200}$ and $\sigma_{cl}$ is expected. 
We compare the dynamical properties of HectoMAP X-ray clusters
 to the theoretical relation given in \citet{Evrard08}
 who derive the relation from the simulated dark matter halos. 
The observed clusters match the model remarkably well. 
\citet{Rines13} argue that this agreement supports the accuracy
 of cluster masses measured from caustic technique.

% ======================================
% Figure 12 - Figure \ref{lxsig}
% ======================================
\begin{figure}
\centering
\includegraphics[scale=0.49]{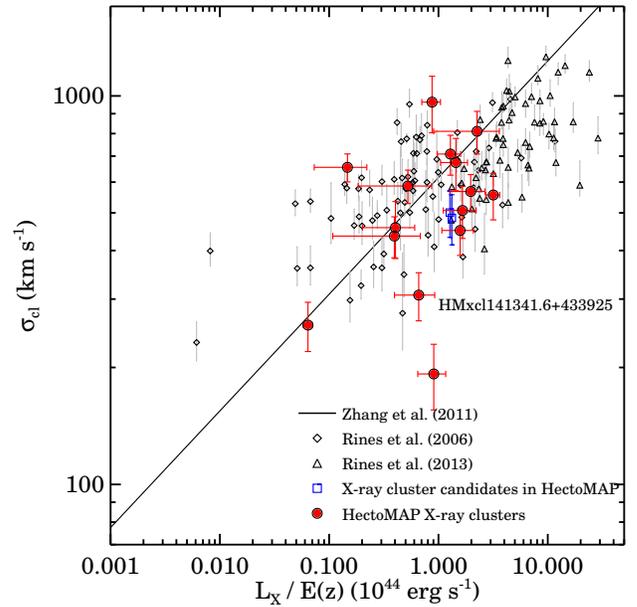}
\caption{
Velocity dispersion ($\sigma_{cl}$) vs. X-ray luminosity ($L_{X}$) for the HectoMAP clusters. 
Red circles show the HectoMAP X-ray clusters and 
 blue squares indicate previously identified X-ray clusters in HectoMAP. 
Gray diamonds and triangles display 
 the CIRS clusters \citep{Rines06} and the HeCS clusters \citep{Rines13},
 respectively. 
The solid line shows the best-fit relation for nearby X-ray cluster samples \citep{Zhang11}. 
Two low $\sigma$ clusters are discussed in Section \ref{scl}. }
\label{lxsig}
\end{figure}
% ======================================

Figure \ref{lxsig} shows 
 the velocity dispersion of the HectoMAP X-ray clusters 
 as a function of their rest-frame X-ray luminosities within the ROSAT band.
For comparison, 
 we plot the CIRS and HeCS clusters 
 that are in a similar redshift and mass range. 

The solid line in Figure \ref{lxsig} shows
 the scaling relation for local clusters from \citet{Zhang11}.
This local scaling relation is consistent with
 the HeCS clusters at a higher redshift range ($0.1 < z < 0.3$). 
The HectoMAP X-ray clusters generally follow 
 the $L_{X} - \sigma_{cl}$ relation 
 defined by previous samples and scaling relations.

We note above that 
 the X-ray emission of HMxcl141341.6+433925 ($z = 0.089$, $\sigma_{cl} \sim 307~\kms$)
 is contaminated by a soft X-ray source. 
Thus, the X-ray flux may be overestimated by a factor of two. 
With this correction, 
 the system moves onto the overall distribution defined by the other systems.

The HectoMAP clusters  
lie on the previously determined scaling relations between CIRS and HeCS samples.
The combination of the sample from the literature
 with the sample we identify 
 represents the first X-ray sample identified with a dense, large area redshift survey to this depth. 
The sample is thus a basis for estimating 
 the number of extended X-ray sources in the RASS data 
 that might be co-identified as candidates with 
 existing photometric surveys like the 
 HSC \citep{Aihara17} and Dark Energy Surveys \citep{DES16} 
 and then tested with spectroscopy.

\subsection{Abundance of X-ray clusters}

The combined spectroscopic and X-ray surveys provide 
 an estimate of the total number of X-ray clusters on the sky. 
To the ROSAT detection limit, 
 $\sim 3 \times 10^{-13}$ erg s$^{-1}$ cm$^{-2}$, 
 we identify 15 clusters within 53 deg$^{2}$
 corresponding to a number density of $\sim 0.3$ deg$^{-2}$. 
Thus, over the entire sky, 
 we expect $\sim12000 \pm 3000$ clusters. 

\citet{Schuecker04} examined the number of galaxy clusters 
 based on the RASS and the SDSS early release data. 
They identified X-ray cluster candidates from the ROSAT X-ray photon map 
 by applying a likelihood function for their cluster search.
They apply a similar likelihood function 
 to the galaxy map from the SDSS photometric catalog 
 for identifying optical cluster candidates.
Then, they cross-matched the X-ray and the optical cluster candidates.
They used SDSS and redshifts in NASA/IPAC Extragalactic Database (NED) to estimate the cluster redshift.
They identified 75 cluster candidates to the X-ray flux limit
 $\sim 3 - 5 \times 10^{-13}$ erg s$^{-1}$ cm$^{-2}$ in the ROSAT energy band $0.1 - 2.4$ keV
 based on a sky coverage of $\sim140$ deg$^{2}$ within $z \leq 0.5$. 
Their estimate suggests that 
 there are $\sim 4000$ X-ray cluster candidates to the X-ray flux limit 
 in the total SDSS sky coverage ($\sim 7000$ deg$^{2}$, for their calculation), 
 yielding $\sim 22000 \pm 2600$ systems in the entire sky. 
 
The XXL cluster survey \citep{Pacaud16} provides another large X-ray cluster sample
 for estimating the number of cluster to a limiting x-ray flux.
The XXL survey is based on deep XMM-Newton data covering a total area of 50 deg$^{2}$.
\citet{Pacaud16} identify 100 X-ray extended sources to an X-ray flux limit
 $3 \times 10^{-14}$ erg s$^{-1}$ cm$^{-2}$, 
 an order of magnitude fainter than the RASS. 
They obtain spectroscopic redshifts of most clusters
 from various spectroscopic observation campaigns. 
The redshift range of the XXL clusters is $z < 1.2$ and 
 the median number of spectroscopic members per XXL cluster is only $\sim 6$.

The XXL cluster sample includes more clusters than the HectoMAP X-ray cluster sample 
 because the X-ray flux limit is deeper. 
When we limit the XXL flux limit to the HectoMAP X-ray limit, 
 there are 22 XXL clusters with $z < 0.4$, 
 comparable with the HectoMAP sample. 
Based on the XXL cluster survey, 
 there would be $\sim 18000 \pm 4000$ X-ray systems over the entire sky to this limit.

The total number of X-ray clusters we predict is 
 marginally consistent with the prediction based on \citet{Pacaud16}, 
 but significantly smaller than the estimate of \citet{Schuecker04}. 
Our cluster identification method differs from these studies. 
We require that the system be identifiable in redshift space. 
This requirement removes superpositions that can masquerade as overdensities on the sky.
Thus we might expect a somewhat smaller number of systems. 
Cosmic variance may contribute to the marginal agreement between HectoMAP and XXL.
Among the three samples, \citet{Pacaud16} have by far the deepest X-ray data. 
The inconsistency between the larger area \citet{Schuecker04} survey and 
 the smaller HectoMAP X-ray and \citet{Pacaud16} samples 
 requires further investigation based on large independent catalogs.

%=============================================================
\section{SUMMARY}\label{summary}

HectoMAP is a dense redshift survey covering the redshift range $z \lesssim 0.7$. 
The survey is sufficiently dense that 
 massive clusters of galaxies can be identified 
 in redshift space throughout this range. 
As a step toward construction of a complete catalog of systems 
 we compare a FoF catalog with the RASS 
 to identify extended X-ray sources. 
We also cross-identify the X-ray systems 
 with the available Hyper Suprime-Cam imaging of the HectoMAP region. 
The images confirm the robustness of the cluster identification.

We identify 15 massive galaxy clusters (7 are new) 
 based on combining HectoMAP with the RASS. 
We apply an FoF algorithm to identify systems in redshift space 
 and cross-identify the ROSAT X-ray extended emission. 
The clusters we identify contain $\gtrsim 20$ spectroscopically identified members.
The cluster survey is complete to the X-ray flux limit of  
 $\sim 3 \times 10^{-13}$ erg s$^{-1}$ cm$^{-2}$. 
We also publish 1036 redshifts for the cluster members.

We also revisit known X-ray cluster candidates from the literature 
 based on the HectoMAP spectroscopic sample and the RASS. 
We find no additional clusters above the flux limit suggesting that 
 our flux-limited sample is complete. 
Among the candidate systems in the literature, 
 four are not confirmed by the spectroscopic data. 
These X-ray sources may be contaminated by background AGNs.
This test underscores 
 the importance of dense spectroscopic samples 
 for identifying galaxy clusters with multi-wavelength data.
 
The HectoMAP X-ray clusters 
 generally follow the scaling relations derived from known massive X-ray clusters: 
 $M_{200} - \sigma_{cl}$ and $L_{X} - \sigma_{cl}$ relations. 
A few poorly sampled systems are outliers. 
 
Our cluster survey predicts
 $\sim12000 \pm 3000$ detectable X-ray clusters in the RASS 
 with $\sim 3 \times 10^{-13}$ erg s$^{-1}$ cm$^{-2}$ and within $z \lesssim 0.4$. 
To the same flux limit, 
 our prediction is consistent with a prediction based on the XXL survey \citep{Pacaud16},
 but is significantly below the prediction by \citet{Schuecker04}.  
The e-ROSITA flux limit should resolve this issue and will enable detection of 
 massive clusters throughout the HectoMAP redshift range 
 along with a much greater cluster mass range at redshifts $\lesssim 0.4$. 
The combination of HectoMAP dense spectroscopy, 
 complete {\it Subaru} imaging of the entire HectoMAP field, and the e-ROSITA survey
 should provide a robust catalog of clusters for
 increasingly sophisticated tests of cluster evolution and
 for determination of the cosmological parameters.

\acknowledgments
We thank an anonymous referee for carefully reading the manuscript that improved the clarity of the paper. 
The referee kindly pointed out a mistake in Table 1. 
J.S. gratefully acknowledges the support of the CfA Fellowship.  
M.J.G is supported by the Smithsonian Institution. 
G.C. and H.B. acknowledge the support by Deutsche Forschungsgemeinschaft through the Transregio project TR33
 and through the Excellence Cluster ``Origin and Evolution of the Universe".
AD acknowledges partial support from the INFN grant InDark.  
We thank Susan Tokarz for reducing the spectroscopic data and 
 Micheal Kurtz, Perry Berlind and Mike Calkins for assisting with the observations. 
We also thank the telescope operators at the MMT and Nelson Caldwell
 for scheduling Hectospec queue observations.
We thank the HSC help desk team, especially Michitaro Koike and Sogo Mineo, for making the useful tools available.
J.S. acknowledges Felipe Andrade-Santos for his help in using PIMMS.
This research has made use of NASA’s Astrophysics Data System Bibliographic Services.  

We have made use of the ROSAT Data Archive of 
 the Max-Planck-Institut f{\"u}r extraterrestrische Physik (MPE) at Garching, Germany.
This research has made use of the X-Rays Clusters Database (BAX)
 which is operated by the Laboratoire d'Astrophysique de Tarbes-Toulouse (LATT),
 under contract with the Centre National d'Etudes Spatiales (CNES).

The Hyper Suprime-Cam (HSC) collaboration includes 
 the astronomical communities of Japan and Taiwan, and Princeton University. 
The HSC instrumentation and software 
 were developed by the National Astronomical Observatory of Japan (NAOJ), 
 the Kavli Institute for the Physics and Mathematics of the Universe (Kavli IPMU),
 the University of Tokyo, the High Energy Accelerator Research Organization (KEK), 
 the Academia Sinica Institute for Astronomy and Astrophysics in Taiwan (ASIAA), 
 and Princeton University. 
Funding was contributed by the FIRST program from Japanese Cabinet Office, 
 the Ministry of Education, Culture, Sports, Science and Technology (MEXT), 
 the Japan Society for the Promotion of Science (JSPS), Japan Science and Technology Agency (JST), 
 the Toray Science Foundation, NAOJ, Kavli IPMU, KEK, ASIAA, and Princeton University. 

This paper makes use of software developed for 
 the Large Synoptic Survey Telescope. 
We thank the LSST Project for making their code available as free software at  http://dm.lsst.org

The Pan-STARRS1 Surveys (PS1) have been made possible through 
 contributions of the Institute for Astronomy, the University of Hawaii, 
 the Pan-STARRS Project Office, the Max-Planck Society and its participating institutes, 
 the Max Planck Institute for Astronomy, Heidelberg and the Max Planck Institute for Extraterrestrial Physics, Garching, 
 the Johns Hopkins University, Durham University, the University of Edinburgh, Queen’s University Belfast, 
 the Harvard-Smithsonian Center for Astrophysics, the Las Cumbres Observatory Global Telescope Network Incorporated, 
 the National Central University of Taiwan, the Space Telescope Science Institute, 
 the National Aeronautics and Space Administration under Grant No. NNX08AR22G issued 
 through the Planetary Science Division of the NASA Science Mission Directorate, 
 the National Science Foundation under Grant No. AST-1238877, the University of Maryland, 
 and Eotvos Lorand University (ELTE) and the Los Alamos National Laboratory.

Based on data collected at the Subaru Telescope and retrieved from the HSC data archive system, 
 which is operated by Subaru Telescope and Astronomy Data Center at National Astronomical Observatory of Japan.
%\clearpage

%=============================================================================================================
%  Bibliograph - References
%=============================================================================================================

\clearpage

%=============================================================================================================
%  Tables 
%=============================================================================================================
%=================================
%%%   Table 1
%=================================
\begin{turnpage}
\begin{deluxetable}{llccccccccc}
%\rotate
\tablecolumns{11}
\tabletypesize{\scriptsize}
\tablewidth{0pt}
\tablecaption{HectoMAP X-ray clusters}
\tablehead{
\colhead{ID} & \colhead{known ID} &
               \colhead{R.A.$_{\rm Caustic}$} & \colhead{Decl.$_{\rm Caustic}$} & \colhead{z} & \colhead{N$_{\rm mem}$} & 
               \colhead{R.A.$_{\rm X-ray}$}   & \colhead{Decl.$_{\rm X-ray}$}   & \colhead{$r_{\rm offset}$\tablenotemark{*}} & \colhead{$f_{X}$\tablenotemark{a}} & \colhead{$L_{X}$\tablenotemark{b}} \\
\colhead{}   & \colhead{}         & 
               \colhead{}                     & \colhead{}                      & \colhead{}  &                         &
               \colhead{}                     & \colhead{}           & \colhead{(arcmin)}     & \colhead{10$^{-13}$ erg cm$^{-2}$ s$^{-1}$} & \colhead{($10^{44}$ erg s$^{-1}$)} } 
\startdata
HMxcl162726.7+424052 &            RXCJ1627.3+4240 & 16:27:26.7 & +42:40:52 & 0.032 &  33 & 16:27:25.0 & 42:40:24 & 0.369 & $27.50 \pm 0.28$ & $0.06 \pm 0.01$ \\
HMxcl141341.6+433925 &                      A1885 & 14:13:41.6 & +43:39:25 & 0.089 &  40 & 14:13:38.7 & 43:40:15 & 1.797 & $34.02 \pm 1.36$ & $0.66 \pm 0.26$ \\
HMxcl162134.3+424558 &                      A2183 & 16:21:34.3 & +42:45:58 & 0.135 &  51 & 16:21:28.0 & 42:45:03 & 0.643 & $19.97 \pm 0.39$ & $0.87 \pm 0.17$ \\
HMxcl151550.0+434556 &            MCXC1515.5+4346 & 15:15:50.0 & +43:45:56 & 0.137 &  18 & 15:15:33.0 & 43:46:35 & 0.417 &  $8.08 \pm 0.40$ & $0.40 \pm 0.20$ \\
HMxcl162632.8+424039 & A2192, WHLJ162642.5+424012 & 16:26:32.8 & +42:40:39 & 0.187 & 110 & 16:26:41.1 & 42:40:13 & 0.263 &  $1.16 \pm 0.06$ & $0.15 \pm 0.07$ \\
HMxcl142837.5+433852 &                         -- & 14:28:37.5 & +43:38:52 & 0.213 &  30 & 14:28:39.3 & 43:40:14 & 1.368 &  $9.79 \pm 0.23$ & $1.29 \pm 0.31$ \\
HMxcl150730.7+424424 &        WHLJ150723.2+424402 & 15:07:30.7 & +42:44:24 & 0.218 & 108 & 15:07:31.7 & 42:44:39 & 1.689 & $13.42 \pm 0.43$ & $1.58 \pm 0.51$ \\
HMxcl163445.9+424641 &                         -- & 16:34:45.9 & +42:46:41 & 0.224 & 218 & 16:35:16.0 & 43:08:23 & 1.687 &  $3.90 \pm 0.25$ & $0.53 \pm 0.34$ \\
HMxcl150859.8+425011 &                         -- & 15:08:59.8 & +42:50:11 & 0.241 &  28 & 15:09:01.3 & 42:49:58 & 0.658 &  $2.17 \pm 0.16$ & $0.40 \pm 0.29$ \\
HMxcl153606.7+432527 &                         -- & 15:36:06.7 & +43:25:27 & 0.255 &  31 & 15:36:02.6 & 43:26:40 & 0.994 & $17.75 \pm 0.26$ & $3.17 \pm 0.46$ \\
HMxcl143543.4+433828 &                         -- & 14:35:43.4 & +43:38:28 & 0.267 &  57 & 14:35:41.7 & 43:36:43 & 0.907 & $10.82 \pm 0.38$ & $1.97 \pm 0.69$ \\
HMxcl163352.9+430529 &        WHLJ163355.8+430528 & 16:33:52.9 & +43:05:29 & 0.271 &  54 & 16:33:48.9 & 43:04:13 & 1.762 &  $4.07 \pm 0.12$ & $0.91 \pm 0.26$ \\
HMxcl141109.9+434145 &        WHLJ141115.4+434123 & 14:11:09.9 & +43:41:45 & 0.299 & 123 & 14:11:12.2 & 43:41:43 & 0.931 &  $5.64 \pm 0.16$ & $1.44 \pm 0.40$ \\
HMxcl145913.1+425808 &        WHLJ145912.8+425758 & 14:59:13.1 & +42:58:08 & 0.371 &  39 & 14:59:08.2 & 42:57:50 & 2.239 &  $2.96 \pm 0.10$ & $1.66 \pm 0.55$ \\
HMxcl132730.5+430433 &                         -- & 13:27:30.5 & +43:04:33 & 0.372 &  99 & 13:27:28.5 & 43:06:03 & 1.038 &  $5.82 \pm 0.35$ & $2.26 \pm 1.35$ 
\enddata
\label{xcl}
\tablenotetext{*}{The distance of the center of X-ray emission from the BCGs. }
\tablenotetext{a}{X-ray flux we measure from ROSAT in units of 10$^{-13}$ erg cm$^{-2}$ s$^{-1}$. 
The flux is obtained from the plateau of the growth curve analysis method within the rest-frame energy band $0.5 - 2.0$ keV. } 
\tablenotetext{b}{X-ray luminosity we measure from ROSAT in units of 10$^{44}$ erg s$^{-1}$. 
The luminosity is corrected to give the values within an aperture of $R_{500}$.} 
\end{deluxetable}
\end{turnpage}

%=================================
%%%   Table 2
%=================================
\begin{deluxetable}{lccc}
\tablecolumns{4}
\tabletypesize{\scriptsize}
\tablewidth{0pt}
\tablecaption{Dynamical Properties of HectoMAP X-ray Clusters}
\tablehead{
\colhead{ID} & \colhead{$R_{200}$} & \colhead{$\sigma_{cl}$\tablenotemark{*}} & \colhead{$M_{200}$} \\
\colhead{}   & \colhead{(Mpc)}     & \colhead{$(\kms)$}                       & \colhead{($10^{14} M_{\odot}$)}}
\startdata
HMxcl162726.7+424052 & $0.498 \pm ^{0.058}_{0.068}$ & $257.1 \pm  36.7$ & $0.145 \pm ^{0.044}_{0.044}$ \\
HMxcl141341.6+433925 & $0.551 \pm ^{0.013}_{0.014}$ & $307.1 \pm  42.9$ & $0.207 \pm ^{0.011}_{0.011}$ \\
HMxcl162134.3+424558 & $1.152 \pm ^{0.099}_{0.120}$ & $963.9 \pm 162.4$ & $1.983 \pm ^{0.555}_{0.555}$ \\
HMxcl151550.0+434556 & $0.823 \pm ^{0.088}_{0.098}$ & $458.1 \pm  76.4$ & $0.722 \pm ^{0.181}_{0.181}$ \\
HMxcl162632.8+424039 & $1.197 \pm ^{0.130}_{0.159}$ & $654.9 \pm  55.0$ & $2.347 \pm ^{0.596}_{0.596}$ \\
HMxcl142837.5+433852 & $1.193 \pm ^{0.011}_{0.012}$ & $708.5 \pm  80.9$ & $2.388 \pm ^{0.069}_{0.069}$ \\
HMxcl150730.7+424424 & $0.898 \pm ^{0.028}_{0.027}$ & $450.5 \pm  63.2$ & $1.024 \pm ^{0.059}_{0.059}$ \\
HMxcl163445.9+424641 & $1.168 \pm ^{0.004}_{0.004}$ & $586.3 \pm  58.0$ & $2.271 \pm ^{0.017}_{0.017}$ \\
HMxcl150859.8+425011 & $0.864 \pm ^{0.069}_{0.077}$ & $435.8 \pm  51.3$ & $0.934 \pm ^{0.168}_{0.168}$ \\
HMxcl153606.7+432527 & $0.986 \pm ^{0.070}_{0.077}$ & $555.9 \pm  81.2$ & $1.413 \pm ^{0.228}_{0.228}$ \\
HMxcl143543.4+433828 & $1.006 \pm ^{0.102}_{0.131}$ & $567.5 \pm  60.9$ & $1.518 \pm ^{0.373}_{0.373}$ \\
HMxcl163352.9+430529 & $0.613 \pm ^{0.023}_{0.023}$ & $192.5 \pm  37.1$ & $0.346 \pm ^{0.025}_{0.025}$ \\
HMxcl141109.9+434145 & $1.311 \pm ^{0.022}_{0.024}$ & $673.8 \pm 107.3$ & $3.485 \pm ^{0.175}_{0.175}$ \\
HMxcl145913.1+425808 & $1.047 \pm ^{0.029}_{0.029}$ & $507.2 \pm  78.6$ & $1.925 \pm ^{0.142}_{0.142}$ \\
HMxcl132730.5+430433 & $1.298 \pm ^{0.088}_{0.099}$ & $810.7 \pm 101.1$ & $3.679 \pm ^{0.688}_{0.688}$
\enddata
\tablenotetext{*}{The error is the $1\sigma$ deviation derived from 1000 time bootstrap resamplings for cluster members within $R_{200}$.}
\label{xcl_dyn}
\end{deluxetable}

%=================================
%%%   Table 3
%=================================
\begin{deluxetable}{lcccccc}
\tablecolumns{7}
\tabletypesize{\footnotesize}
\tablewidth{0pt}
\tablecaption{Members of HectoMAP X-ray clusters} 
\tablehead{
\colhead{Cluster ID} & \colhead{SDSS Object ID} & \colhead{R.A.} & \colhead{Decl.} & \colhead{z} & \colhead{$z_{err}$} & \colhead{z Source}}
\startdata
HMxcl162726.7+424052 & 1237655348358480098 & 246.267148 & +42.509578 & 0.03166 & 0.00001 & SDSS \\
HMxcl162726.7+424052 & 1237655473430135086 & 246.323461 & +42.694233 & 0.03148 & 0.00002 & SDSS \\
HMxcl162726.7+424052 & 1237655473430200607 & 246.571651 & +42.673512 & 0.03168 & 0.00016 &  MMT \\
HMxcl162726.7+424052 & 1237655348895285690 & 246.712111 & +42.826501 & 0.03203 & 0.00003 &  MMT \\
HMxcl162726.7+424052 & 1237655348895351148 & 246.866560 & +42.806850 & 0.03002 & 0.00002 & SDSS \\
HMxcl162726.7+424052 & 1237655348895416661 & 246.971122 & +42.652934 & 0.03154 & 0.00002 & SDSS \\
HMxcl162726.7+424052 & 1237655348895481992 & 246.958335 & +42.563954 & 0.03148 & 0.00013 &  MMT \\
HMxcl162726.7+424052 & 1237655473430331641 & 246.855410 & +42.514434 & 0.03146 & 0.00001 & SDSS \\
HMxcl162726.7+424052 & 1237655473967136890 & 247.216951 & +42.812006 & 0.03158 & 0.00001 & SDSS \\
HMxcl162726.7+424052 & 1237655348895351215 & 246.823539 & +42.695248 & 0.03140 & 0.00001 & SDSS
\enddata
\label{xclmem}
\tablecomments{
A portion of the table is shown for guidance regarding its format.
The entire table is available in machine-readable form in the online journal. }
\end{deluxetable}
\clearpage

%=================================
%%%   Table 4
%=================================
\begin{turnpage}
\begin{deluxetable}{lcccccccccccc}
%\rotate
\tablecolumns{13}
\tabletypesize{\scriptsize}
\tablewidth{0pt}
\tablecaption{HectoMAP X-ray clusters from the literature}
\tablehead{
\colhead{ID} & 
\colhead{R.A.$_{\rm cat}$}        & \colhead{Decl.$_{\rm cat}$}       & \colhead{z$_{\rm cat}$}     & 
\colhead{R.A.$_{\rm caustic}$}    & \colhead{Decl.$_{\rm caustic}$}   & \colhead{z$_{\rm caustic}$} & \colhead{N$_{mem}$} &
\colhead{$f_{\rm X, lit}$\tablenotemark{a}}   & \colhead{$L_{\rm X, lit}$\tablenotemark{b}}   & 
\colhead{$f_{\rm X, ROSAT}$\tablenotemark{c}} & \colhead{$L_{\rm X, ROSAT}$\tablenotemark{d}} & \colhead{ref.\tablenotemark{*}} }
\startdata
 RXCJ1627.3+4240$^{\star}$   & 16:27:23.6 & +42:40:42.0 & 0.0317 & 16:27:26.7 & 42:40:52.6 & 0.0314  &  33     & 27.50 & 0.06 & $27.50 \pm 0.28$ & $0.06 \pm 0.01$ &  1 \\
       ABELL1885$^{\star}$   & 14:13:46.7 & +43:40:01.6 & 0.0890 & 14:13:43.4 & 43:39:48.2 & 0.0888  &  49     & 52.00 & 1.02 & $34.02 \pm 1.36$ & $0.66 \pm 0.26$ &  1, 8 \\
MCXCJ1515.5+4346$^{\star}$   & 15:15:32.9 & +43:46:35.0 & 0.1370 & 15:15:49.8 & 43:57:25.9 & 0.1342  &  18     &  8.03 & 0.38 & $8.08 \pm 0.40$ & $0.40 \pm 0.20$  &  3, 6 \\
       ABELL2192$^{\star}$   & 16:26:37.2 & +42:40:19.7 & 0.1880 & 16:26:40.5 & 42:39:26.0 & 0.1874  &  61     & 27.50 & 2.73 & $1.16 \pm 0.06$ & $0.15 \pm 0.07$  &  1, 8 \\
J150723.2+424402$^{\star}$   & 15:07:23.2 & +42:44:02.8 & 0.2173 & 15:07:30.7 & 42:44:17.5 & 0.2184  & 106     &  6.54 & 0.92 & $13.42 \pm 0.43$ & $1.58 \pm 0.51$ &  8 \\
J152347.2+434945 		     & 15:23:47.2 & +43:49:45.9 & 0.2189 & \nodata    & \nodata    & \nodata & \nodata &  1.47 & 0.21 & $<2.0$  &   0.2814 &  8 \\
MCXCJ1515.6+4350 		     & 15:15:36.8 & +43:50:50.0 & 0.2430 & 15:15:15.8 & 43:39:55.2 & 0.2414  & 134     &  6.11 & 1.07 & \nodata &  \nodata &  3, 4 \\
J163355.8+430528$^{\star}$   & 16:33:55.8 & +43:05:28.2 & 0.2699 & 16:33:54.9 & 43:05:45.0 & 0.2705  &  49     &  1.35 & 0.31 & \nodata &  \nodata &  8 \\
J134117.1+431126$^{\rm e}$   & 13:41:17.1 & +43:11:26.8 & 0.2725 & 13:41:16.3 & 43:11:25.1 & 0.2725  &  24     &  5.73 & 1.33 & $<2.7$ &   0.6243 &  3, 8 \\
       ABELL2198$^{\rm f}$   & 16:28:04.7 & +43:49:25.7 & 0.0800 & 16:28:14.0 & 43:48:57.3 & 0.2767  &  37     &  5.31 & 1.27 & $<3.0$ &   0.7183 &  7 \\
J141115.4+434123$^{\star}$   & 14:11:15.4 & +43:41:24.0 & 0.2980 & 14:11:09.9 & 43:41:45.6 & 0.2998  & 111     &  2.21 & 0.64 & $5.64 \pm 0.16$ & $1.44 \pm 0.40$ &  8 \\
       VMF98 163 		     & 14:29:38.1 & +42:34:25.0 & 0.3000 & 14:30:05.6 & 42:51:11.4 & 0.2637  &  47     &  1.84 & 0.39 & $<2.0$  &   0.4290 &  3 \\
J145912.8+425758$^{\star}$   & 14:59:12.8 & +42:57:58.1 & 0.3697 & 14:59:12.0 & 42:58:08.5 & 0.3703  &  34     &  2.26 & 1.06 & $2.96 \pm 0.10$ & $1.66 \pm 0.55$ &  8 \\
  GHO1602+4312$^{\dagger}$   & 16:04:25.2 & +43:04:52.7 & 0.8950 & \nodata    & \nodata    & \nodata & \nodata &  1.86 & 7.38 & \nodata &  \nodata &  5 \\
MCXCJ1429.0+4241$^{\dagger}$ & 14:29:05.8 & +42:41:12.0 & 0.9200 & \nodata    & \nodata    & \nodata & \nodata &  0.91 & 3.87 & \nodata &  \nodata &  2
\enddata
\label{pxcl}
\tablenotetext{*}{(1) \citet{Bohringer00}, (2) \citet{Horner08}, 
(3) \citet{Burenin07}, (4) \citet{Vikhlinin98}, 
(5) \citet{Lubin04}, (6) \citet{Voevodkin10},
(7) \citet{David99}, (8) \citet{Wen09} }
\tablenotetext{$\bigstar$}{Cluster candidates overlapped with the HectoMAP X-ray clusters. }
\tablenotetext{$\dagger$}{Cluster candidates beyond the redshift range of HectoMAP. }
\tablenotetext{a}{X-ray flux from the literature in a unit of 10$^{-13}$ erg cm$^{-2}$ s$^{-1}$}
\tablenotetext{b}{X-ray luminosity from the literature in a unit of 10$^{44}$ erg s$^{-1}$}
\tablenotetext{c}{X-ray flux we measure from ROSAT in a unit of 10$^{-13}$ erg cm$^{-2}$ s$^{-1}$.}
\tablenotetext{d}{X-ray luminosity we measure from ROSAT in a unit of 10$^{44}$ erg s$^{-1}$. }
\tablenotetext{e}{$R_{200} = 0.943 \pm 0.006$ Mpc, $\sigma_{cl} = 485.2 \pm 67.9~\kms$, and $M_{200} = 1.248 \pm 0.020 \times 10^{14}$ M$_{\odot}$}
\tablenotetext{f}{$R_{200} = 0.827 \pm 0.038$ Mpc, $\sigma_{cl} = 500.9 \pm 70.1~\kms$, and $M_{200} = 0.853 \pm 0.094 \times 10^{14}$ M$_{\odot}$}
\end{deluxetable}
\end{turnpage}
\clearpage
\end{document}